\documentclass[prd,tightenlines,floatfix,showpacs,preprintnumbers,nofootinbib,eqsecnum,superscriptaddress,twocolumn]{revtex4-1}

\usepackage{color}      
\usepackage{amsbsy}     
\usepackage{amsfonts}       
\usepackage{amsmath}        
\usepackage{amssymb}        
\usepackage{xcolor}
\usepackage{wasysym}
\usepackage{multirow}
\usepackage{mciteplus}
\usepackage{psfrag}
\usepackage{slashed}
\usepackage{cancel}
\usepackage{graphicx}
\usepackage{ulem}

\newcommand{\DMNLO}{{\tt DM@NLO}}
\newcommand{\MO}{{\tt micrOMEGAs}}

\newcommand{\CHep}{{\tt CalcHEP}}

\newcommand{\SPheno}{{\tt SPheno}}

\newcommand{\DRbar}{{$\overline{\mathrm{DR}}$}}

\newcommand{\beq}{\begin{equation}}
\newcommand{\eeq}{\end{equation}}
\newcommand{\bea}{\begin{eqnarray}}
\newcommand{\eea}{\end{eqnarray}}

\catcode`\@=11
\font\manfnt=manfnt
\def\Watchout{\@ifnextchar [{\W@tchout}{\W@tchout[1]}}
\def\W@tchout[#1]{{\manfnt\@tempcnta#1\relax%
  \@whilenum\@tempcnta>\z@\do{%
    \char"7F\hskip 0.3em\advance\@tempcnta\m@ne}}}
\let\foo\W@tchout
\def\dubious{\@ifnextchar[{\@dubious}{\@dubious[1]}}

\def\@dubious[#1]{%
  \color{red}\setbox\@tempboxa\hbox{\@W@tchout#1}
  \@tempdima\wd\@tempboxa
  \list{}{\leftmargin\@tempdima}\item[\hbox to 0pt{\hss\@W@tchout#1}]}
\def\@W@tchout#1{\W@tchout[#1]}
\catcode`\@=12

\begin{document}
\preprint{LAPTH-035/15, MS-TP-16-03}

\title{Theoretical uncertainty of the supersymmetric dark matter relic density \\ from scheme and
scale variations}


\author{J.~Harz}
 \email{jharz@lpthe.jussieu.fr}
 \affiliation{
	Sorbonne Universit\'es, Institut Lagrange de Paris (ILP), 98\,bis Boulevard Arago, F-75014 Paris, France \\
	Sorbonne Universit\'es, UPMC Univ Paris 06, UMR 7589, LPTHE, F-75005 Paris, France\\
	CNRS, UMR 7589, LPTHE, F-75005 Paris, France 
  }
  
\author{B.~Herrmann}
 \email{herrmann@lapth.cnrs.fr}
 \affiliation{
	LAPTh, Universit\'e Savoie Mont Blanc, CNRS, 9 Chemin de Bellevue, B.P.~110, F-74941 Annecy-le-Vieux, France
  }

\author{M.~Klasen}
 \email{michael.klasen@uni-muenster.de}
 \affiliation{
	Institut f\"ur Theoretische Physik, Westf\"alische Wilhelms-Universit\"at M\"unster, Wilhelm-Klemm-Stra{\ss}e 9, D-48149 M\"unster, Germany
  }

\author{K.~Kova\v{r}\'ik}
 \email{karol.kovarik@uni-muenster.de}
 \affiliation{
	Institut f\"ur Theoretische Physik, Westf\"alische Wilhelms-Universit\"at M\"unster, Wilhelm-Klemm-Stra{\ss}e 9, D-48149 M\"unster, Germany
  }
 

\author{P.~Steppeler}
 \email{p\_step04@uni-muenster.de}
 \affiliation{
	Institut f\"ur Theoretische Physik, Westf\"alische Wilhelms-Universit\"at M\"unster, Wilhelm-Klemm-Stra{\ss}e 9, D-48149 M\"unster, Germany
  }

\date{\today}

\begin{abstract}
For particle physics observables at colliders such as the LHC at CERN, it has been common
practice for many decades to estimate the theoretical uncertainty by studying the variations
of the predicted cross sections with a priori unpredictable scales. In astroparticle physics,
this has so far not been possible, since most of the observables were calculated at Born level
only, so that the renormalization scheme and scale dependence could not be studied in a
meaningful way.
In this paper, we present the first quantitative study of the theoretical uncertainty of the
neutralino dark matter relic density from scheme and scale variations. We first explain
in detail how the renormalization scale enters the tree-level calculations through coupling
constants, masses and mixing angles. We then demonstrate a reduction of the renormalization
scale dependence through one-loop SUSY-QCD corrections in many different dark matter annihilation
channels and enhanced perturbative stability of a mixed on-shell/$\overline{\rm DR}$
renormalization scheme over a pure $\overline{\rm DR}$ scheme in the top-quark sector. In the
stop-stop annihilation channel, the Sommerfeld enhancement and its scale dependence are shown to
be of particular importance. Finally, the impact of our higher-order SUSY-QCD corrections and
their scale uncertainties are studied in three typical scenarios of the phenomenological Minimal
Supersymmetric Standard Model with eleven parameters (pMSSM-11). We find that the theoretical
uncertainty is reduced in many cases and can become comparable to the size of the experimental
one in some scenarios.
\end{abstract}

\pacs{12.38.Bx,12.60.Jv,95.30.Cq,95.35.+d}

\maketitle

\section{Introduction}
\label{Intro}








Eighty years after Zwicky's discovery of dark matter in the Coma galaxy cluster
\cite{Zwicky:1933gu}, its existence in the Universe is now well established from astronomical
observations on many different length scales, but its nature is still unknown
\cite{Klasen:2015uma}. Nevertheless,
precision measurements of the temperature fluctuations of the cosmic microwave background with
the Planck satellite \cite{Ade:2015xua}, supplemented by WMAP polarization data
\cite{Bennett:2012zja}, allow to deduce the global cold dark matter relic density in the Universe
with previously unparalleled precision to
\begin{equation}
 \Omega_{\mathrm{CDM}}h^2 = 0.1199 \pm 0.0022.
 \label{Planck}
\end{equation}
Here, $h$ denotes the present Hubble expansion rate in units of 100 km s$^{-1}$ Mpc$^{-1}$.

An intriguing possibility for the nature of dark matter is a weakly interacting massive particle
(WIMP), as it automatically leads to the correct relic abundance. In the absence of a suitable
WIMP candidate within the Standard Model (SM), the arguably most popular candidate is a light
supersymmetric (SUSY) particle, since SUSY has many other theoretical motivations ranging from
a symmetry between fermions and bosons over a maximal extension of spacetime symmetry to a
stabilization of the Higgs boson mass. In the case of the Minimal Supersymmetric SM (MSSM), the
lightest SUSY particle (LSP) is typically the lightest neutralino, $\tilde{\chi}_1^0$, a mixture
of the fermionic partners of the neutral electroweak gauge and CP-even Higgs bosons.

For a given set of parameters, the current neutralino relic density can be predicted by solving
the Boltzmann equation describing the time evolution of its number density $n_\chi$,
\begin{equation}
 \frac{\mathrm{d}n_\chi}{\mathrm{d}t} = -3 H n_\chi 
 - \left\langle\sigma_{\mathrm{ann}}v\right\rangle \Big[ n_\chi^2 
 - \left( n_\chi^{\mathrm{eq}} \right)^2 \Big].
 \label{Boltzmann1}
\end{equation}
This nonlinear differential equation contains a term proportional to the Hubble expansion
parameter $H$, which is responsible for the expansion of the Universe and thus for the dilution of
matter. Particle physics enters through the second term on the right hand side, which is
proportional to the thermally averaged annihilation cross section $\left\langle\sigma_{\mathrm{ann}}
v\right\rangle$ and describes the annihilation and creation of neutralinos. This thermally
averaged annihilation cross section includes all possible annihilation and coannihilation
channels of neutralinos and other SUSY particles into SM particles.

After solving the Boltzmann equation numerically, the neutralino relic density is obtained via
\begin{equation}
 \Omega_\chi h^2 ~=~ \frac{m_\chi n_\chi}{\rho_{\mathrm{crit}}},
\end{equation}
where $\rho_{\mathrm{crit}}$ is the critical density of the Universe. This value may then be
compared to the very precise measurement given in Eq.\ \eqref{Planck} and allows the restriction
of the MSSM parameter space. More details on the Boltzmann equation and its numerical treatment
can be found in Refs.\ \cite{Gondolo:1990dk,Griest:1990kh,Edsjo:1997bg,Harz:2012fz,Harz:2014gaa}.

During the last years, considerable progress has been made in the calculation of the neutralino
relic density and other dark matter observables with respect to radiative corrections
\cite{Baro:2007em,Bringmann:2013oja,Beneke:2012tg,Hisano:2015rsa,Drees:2009gt}.
More precisely, in recent publications we have
calculated the full $\mathcal{O}(\alpha_s)$ corrections to general gaugino (co)annihilations into
quarks \cite{Herrmann:2007ku,Herrmann:2009wk,Herrmann:2009mp,Herrmann:2014kma}, to neutralino-stop
coannihilation \cite{Harz:2012fz, Harz:2014tma}, and to stop-antistop annihilation into
electroweak final states \cite{Harz:2014gaa}. It is well known that such fixed-order calculations
introduce an artificial (i.e.\ non-physical) dependence on the renormalization scheme and scale.
For good reasons (preservation of gauge and Lorentz invariance), most perturbative calculations,
including ours, employ dimensional regularization or (for the preservervation of SUSY invariance)
dimensional reduction to regularize ultraviolet (and infrared) divergences. 
 A variation of the renormalization scale leads then to a variation of the
(co)annihilation cross sections and thus the predicted value for the relic density. For a
calculation at $\mathcal{O}(\alpha_s)$, the dependence on the scale is of $\mathcal{O}
(\alpha_s^2)$ and thus allows an estimate of the higher-order theoretical uncertainty.

In the present paper we analyze the renormalization scheme and scale dependence of the
supersymmetric dark matter relic density in detail. With the help of three typical
scenarios, covering all three classes of (co)annihilation channels mentioned above, we
explain where and how the renormalization scale enters the calculations already at tree level
and how its influence is reduced by higher-order SUSY-QCD corrections. This allows us
to estimate for the first time the theoretical error on the calculation of the (co)annihilation cross
sections and the neutralino relic density.

This paper is organized as follows: Sec.\ \ref{Technical} is devoted to the discussion of
technical aspects. We define a phenomenological MSSM with eleven free parameters (pMSSM-11), a
mixed on-shell and $\overline{\rm DR}$ renormalization scheme, and analyze the origin and
influence of the renormalization scale $\mu_R$ on our calculations at leading order (LO) and
next-to-leading order (NLO). Furthermore, we discuss our treatment of Coulomb corrections and how
we define and vary the associated Coulomb scale $\mu_C$ in our numerical analysis. In Sec.\
\ref{Numerics}, we then present numerical results for the various (co)annihilation cross
sections and the predicted neutralino relic density for different choices of the renormalization
scheme and scale in three pMSSM-11 benchmark scenarios. Finally, we conclude in Sec.\
\ref{Conclusion}.

\section{Setup and treatment of scales}
\label{Technical}

We first discuss the setup of our numerical calculations, starting from the input
parameters in the phenomenological MSSM (pMSSM), up to the calculation of the
annihilation cross section $\sigma_{\rm ann}$ and the prediction of the relic density
$\Omega_{\tilde{\chi}}h^2$.


As in our previous publications, we work in a pMSSM with eleven parameters, denoted as pMSSM-11
in the following and defined in the $\overline{\rm DR}$ scheme at the scale $\tilde{M} = 1$ TeV
according to the SPA convention \cite{AguilarSaavedra:2005pw}. We identify this scale with our central
renormalization scale $\mu_R^{\rm central}$.
In particular, we parametrize the Higgs sector by the higgsino mass parameter $\mu$, the
ratio of the vacuum expectation values of the two Higgs doublets $\tan\beta$, and the pole
mass of the pseudoscalar Higgs boson $m_A$. The gaugino sector is fixed by the bino ($M_1$),
wino ($M_2$) and gluino ($M_3$) mass parameters, which in our set-up are not related through
any assumptions stemming from Grand Unified Theories. We define a common soft SUSY-breaking
mass parameter $M_{\tilde{q}_{1,2}}$ for the first- and second-generation squarks. The
third-generation squark masses are controlled by the parameter $M_{\tilde{q}_3}$ associated
with sbottoms and left-handed stops and by the parameter $M_{\tilde{u}_3}$ for right-handed
stops. The trilinear coupling in the stop sector is given by $A_t$, while the trilinear
couplings of the other sectors, including $A_b$, are set to zero. Since the slepton sector
is not at the center of our attention, it is parametrized by a single soft parameter
$M_{\tilde{\ell}}$.


With this set of free parameters, we use the numerical program {\tt SPheno 3.3.3} \cite{Porod:2003um}
to obtain the associated physical mass spectrum at the scale $\tilde{M}$. In addition,
{\tt SPheno} is used to evolve all running parameters via the implemented renormalization
group equations (RGEs) from the input scale $\tilde{M}\equiv\mu_R^{\rm central}$ to the
renormalization scale $\mu_R$, where we want to evaluate the (co)annihilation cross sections.
The results from other SUSY spectrum generators can differ, in particular in regions with
important stop coannihilation, but these differences have been reduced by consistent
implementations of two-loop RGEs \cite{Belanger:2005jk}.
Apart from the implicit dependence in some of the physical parameters to be discussed next,
the renormalization scale also appears explicitly in our higher-order QCD calculations in the
form of loop logarithms.


Our QCD calculations at next-to-leading order (NLO) and beyond are performed within a hybrid
on-shell/$\overline{\rm DR}$ renormalization scheme, described in detail in Refs.\
\cite{Herrmann:2014kma, Harz:2012fz, Harz:2014tma}.
In the quark sector, the top- and bottom-quark masses are defined on-shell and in the
$\overline{\rm DR}$ scheme, respectively, while all other quarks are taken as massless.
Note that through the Yukawa coupling to (in particular the neutral pseudoscalar) Higgs-boson
resonances, the bottom-quark mass can have a sizeable influence on the dark matter annihilation
cross section and must therefore be treated with particular care. We obtain it from the
SM $\overline{\rm MS}$ mass $m_b(m_b)$, determined in an analysis of $\Upsilon$ sum rules,
through evolution to the scale $\mu_R$, transformation to the SM $\overline{\rm DR}$ and
then MSSM $\overline{\rm DR}$ scheme \cite{Herrmann:2014kma, Harz:2012fz}.
In the squark sector, we have five independent parameters 
\begin{equation}
 m_{\tilde{t}_1}, \quad m_{\tilde{b}_1}, \quad m_{\tilde{b}_2}, \quad A_t \quad\mathrm{and}\quad A_b=0.
 \label{eq:RenInput}
\end{equation}
This choice makes our renormalization scheme applicable to all annihilation and coannihilation
processes, where squarks play an important role.
The lighter stop mass and the two sbottom masses are taken to be on-shell, while the stop and 
sbottom trilinear coupling parameters are taken in the $\overline{\rm DR}$ scheme. From these
parameters, we compute as dependent quantities the stop and sbottom mixing angles
$\theta_{\tilde{t}}$ and $\theta_{\tilde{b}}$ in the $\overline{\rm DR}$ scheme at the scale
$\mu_R$ as well as the on-shell masses of the first- and second generation squarks and
$m_{\tilde{t}_2}$ for the heavier stop \cite{Harz:2012fz}.
The strong coupling constant $\alpha_s(\mu_R)$ is renormalized in the MSSM $\overline{\rm DR}$
scheme with six active flavors and obtained after evolution of the world-average, five-flavor
SM $\overline{\rm MS}$ value at the $Z^0$-boson mass to the renormalization scale $\mu_R$ and
an intermediate transformation to the SM $\overline{\rm DR}$ scheme \cite{Harz:2014tma}.
In our calculaton, all running parameters are thus evaluated at the scale $\mu_R$, which
we can then vary in order to study the scale dependence of the (co)annihilation cross sections
and consequently the neutralino relic density. In the following numerical analysis, we will
vary the renormalization scale between $\mu_R = 0.5$ TeV and $\mu_R = 2$ TeV, corresponding to
a factor $1/2$ or $2$, respectively, as compared to the central scale $\mu_R^{\rm central}$.
Note that since we are studying QCD corrections, we are not interested in the effects from the
running of electroweak parameters. We therefore read all SUSY parameters except $A_t$ and $A_b$
from the spectrum obtained at the central scale. 


In our calculation of stop-antistop annihilation into electroweak (EW) final states, we also
include the Coulomb corrections connected to the exchange of potential gluons in the initial
state. As outlined in Ref.\ \cite{Harz:2014gaa}, the exchange of $n$ potential gluons comes with a
factor $(\alpha_s/v)^n$. In the region of low relative velocity $v\lesssim \alpha_s$ between
the incoming stop-antistop pair, this factor can become large and spoil the convergence of the
perturbative series in $\alpha_s$. Hence, these multiple gluon exchanges have to be resummed to
all  orders in perturbation theory in order to get a reliable result. 
Following Ref.\ \cite{Kiyo:2008bv}, this can be done via
\begin{align}
 \label{Coulomb_sigma}
 &\sigma^{\mathrm{Coul}} \big(\tilde{t}_1 \tilde{t}^*_1 \rightarrow \mathrm{EW} \big)  = \frac{4\pi}{v m_{\tilde{t}_1}^2}\\  
 &\times \Im \Big\{ G^{[1]} \big(\mathbf{r}=0;\sqrt{s}+i\Gamma_{\tilde{t}_1},\mu_C \big) \Big\} 
 \sigma^{\mathrm{LO}} \big(\tilde{t}_1\tilde{t}^*_1 \rightarrow \mathrm{EW} \big). \nonumber
\end{align}


As apparent from Eq.\ (\ref{Coulomb_sigma}), we encounter a residual scale dependence of our
final result $\sigma^{\mathrm{Coul}}$ on the Coulomb scale $\mu_C$ connected to the dimensional
regularization of the associated color-singlet Greens function $G^{[1]}$, which is UV-divergent
at the origin $\mathbf{r}=0$ \cite{Beneke:1999zr}. The latter can be obtained as a solution to
the would-be stoponium Schr\"odinger equation
\begin{align}
 \label{Schr-Eq}
 \Big[ H^{[1]} - \big( \sqrt{s}+i\Gamma_{\tilde{t}_1} \big) \Big]
 G^{[1]} \big( \mathbf{r};\sqrt{s}+i\Gamma_{\tilde{t}_1},\mu_C\big)
 = \delta^{(3)}(\mathbf{r}) 
\end{align}
with the Hamilton operator 
\begin{align}
 H^{[1]} = -\frac{1}{m_{\tilde{t}_1}}\Delta + 2 m_{\tilde{t}_1}+V^{[1]}(\mathbf{r})
\end{align}
and the color-singlet Coulomb potential $V^{[1]}(\mathbf{r})$. At NLO, $V^{[1]}(\mathbf{r})$
includes corrections up to $\mathcal{O}(\alpha_s^2)$. Following the results given in Ref.\
\cite{Harz:2014gaa}, Eq.\ (\ref{Coulomb_sigma}) adds up terms of the order $\mathcal{O}((\alpha_s/v)^n)$
and $\mathcal{O}((\alpha_s^2/v))^n)$ up to infinite order in $n$. At this precision $\mu_C$ can be
treated independently from the renormalization scale $\mu_R$ connected to the hard part of the
process \cite{Kiyo:2008bv}. We choose 
\begin{equation}
 \label{Coulomb-scale}
 \mu_C^{\rm central}=\max\{\mu_B,2m_{\tilde{t}_1}v\} 
\end{equation}
as our central
Coulomb scale in order to minimize potentially large logarithms \cite{Beneke:2010da}.
Here, $\mu_B=C_F \alpha_s m_{\tilde{t}_1}$ corresponds to the inverse Bohr radius, the
characteristic energy scale of the stop-antistop bound state. We describe in detail how we
combine the uncertainties due to variations of $\mu_R$ and $\mu_C$ in Sec.\ \ref{QQChapter}.


For values of $\mu_C$ below $\mu_B$, the authors of Ref.\ \cite{Beneke:2005hg} encountered an
instability of the convergence of the perturbative solution used here against an exact numerical
solution of Eq.\ (\ref{Schr-Eq}). This instability has been traced back to large higher-order
logarithmic corrections which tend to blow up for $\mu_C\ll \mu_B$. For the scenarios presented
here we have explictly checked that the variation does not become too large and, hence, does not
yield an overestimation of the corresponding scale uncertainty. This should be kept in mind for
the results shown further below. Finally note that the solution $G^{[1]}$ to Eq.\ (\ref{Schr-Eq})
expanded up to $\mathcal{O}(\alpha_s)$ takes the form 
\begin{align}
 \label{Greensfunc-expand}
 &\Im \Big\{ G^{[1]} \big(0;\sqrt{s} + i \Gamma_{\tilde{t}_1},\mu_C \big) \Big\} = \\ 
 & m_{\tilde{t}_1}^2 \Im \Big\{ \frac{v}{4\pi} \Big[ i + \frac{\alpha_s(\mu_C) C_F}{v}
 \Big( \frac{i \pi}{2} + \ln \frac{\mu_C}{2m_{\tilde{t}_1}v} \Big)
 + \mathcal{O}(\alpha_\mathrm{s}^2) \Big] \Big\}.  \nonumber
\end{align}
Hence, the explicit scale dependence of the Greens function drops out of the Coulomb corrected
cross section at $\mathcal{O}(\alpha_s)$ and re-enters first at NNLO. However, whereas
$m_{\tilde{t}_1}$, which we choose to renormalize in the on-shell scheme, remains scale independent,
Eq.\ (\ref{Greensfunc-expand}) still features an implicit dependence on $\mu_C$ due to the strong
coupling $\alpha_s(\mu_C)$.


As a final remark, let us stress again that, although the renormalization scale appears
explicitly only at the one-loop level, also the tree-level result will vary with the scale
$\mu_R$. First, this is due to taking the values of $A_t$ and $A_b$ at this scale and deriving
from them, e.g., the squark mixing angles, which enter the calculation already at the tree
level whenever top or bottom squarks are involved. As already mentioned, also the bottom-quark
mass and the strong coupling constant are taken as MSSM \DRbar\ parameters and thus directly
scale-dependent. The former appears at the tree level in Yukawa couplings, e.g., in the
Higgs funnel of gaugino annihilation into bottom quark pairs, and the latter enters at tree
level in stop coannihalation into top quarks and gluons, as we explain in more detail in the
next Section.

\section{Numerical results}
\label{Numerics}


\begin{table*}
 \caption{pMSSM-11 input parameters for three selected reference scenarios. All parameters
          except $\tan\beta$ are given in GeV. }
 \label{ScenarioList}
 \begin{tabular}{|c|ccccccccccc|}
 \hline
 $\quad$ & $\quad\tan\beta\quad$ & $\quad\mu\quad$ & $\quad m_A\quad$ & $\quad M_1\quad$ & $\quad M_2\quad$ & $\quad M_3\quad$ & 		$\quad M_{\tilde{q}_{1,2}}\quad$ & $\quad M_{\tilde{q}_3}\quad$ & $\quad M_{\tilde{u}_3}\quad$ & $\quad M_{\tilde{\ell}}\quad$& $\quad A_t\quad$ \\ 
 \hline 
 A & 13.4 & 1286.3 & 1592.9 & 731.0 & 766.0 & 1906.3 & 3252.6 & 1634.3 & 1054.4 & 3589.6 & -2792.3\\			
 B & 27.0 & 2650.8 & 1441.5 & 1300.0 & 1798.4 & 1744.8 & 2189.7 & 2095.3 & 1388.0 & 1815.5 & -4917.5\\	
 C &  5.8 & 2925.8 & 948.8 & 335.0 & 1954.1 & 1945.6 & 3215.1 & 1578.0 & 609.2 & 3263.9 & 3033.7\\			
 \hline
 \end{tabular}
\end{table*}

\begin{table*}
 \caption{Gaugino and stop masses, the decomposition of the lightest neutralino, and selected
          observables corresponding to the reference scenarios of Tab.\ \ref{ScenarioList}.
          All masses are given in GeV.}
 \label{ScenarioProps}
 \begin{tabular}{|c|cc|cc|cc|cccc|ccc|}
 \hline
 $\quad$  & ~~ $m_{\tilde{\chi}^0_1}$~~ & ~~$m_{\tilde{\chi}^0_2}$~~ &  ~~$m_{\tilde{\chi}^{\pm}_1}$~~ & ~~$m_{\tilde{\chi}^{\pm}_2}$~~ & ~~$m_{\tilde{t}_1}$~~ & ~~$m_{\tilde{t}_2}$~~ & ~~$Z_{1\tilde{B}}$~~ &  ~~$Z_{1\tilde{W}}$~~ &  ~~$Z_{1\tilde{H}_1}$~~ &  ~~$Z_{1\tilde{H}_2}$~~ & ~~$m_{h^0}$~~ & ~~$\Omega_{\tilde{\chi}^0_1} h^2$~~ & $\mathrm{BR}(b\rightarrow s\gamma)$ \\
 \hline 
 A & 738.1 & 802.5 & 802.4 & 1295.3 & 1032.1 & 1682.0 & - 0.996 & 0.049 & -0.059 & 0.037 & 126.5 & 0.1248 & $3.0\cdot 10^{-4}$ \\			
 B &1306.3 & 1827.0 & 1827.2 & 2640.0 & 1361.7 & 2157.3 & -1.000 & 0.002 & -0.024 &  0.013 & 123.7 & 0.1134 &  $3.1\cdot 10^{-4}$\\
 C & 338.3& 1996.6 & 1996.7 & 2909.0& 376.3& 1554.0 & 1.000& 0.000& 0.016 & -0.004 & 121.7 & 0.1193& $3.49\cdot 10^{-4}$ \\	
 \hline		
 \end{tabular}
\end{table*}

\begin{table}
 \caption{Most relevant (co)annihilation channels in the reference scenarios of Tab.\
          \ref{ScenarioList}. Channels which contribute less than 1\% to the thermally averaged
          cross section and/or are not implemented in our code are not shown.}
 \label{ScenarioChannels}
 \begin{tabular}{|rl|cccc|}
 \hline
 &  & ~~~~ A ~~~~ & ~~~~ B ~~~~ & ~~~~ C ~~~~ &  \\
 \hline
 $\tilde{\chi}^0_1 \tilde{\chi}^0_1 \to$ & $t\bar{t}$ & 2\% & & 16\% & \\
                                        & $b\bar{b}$ & 9\% &  &  & \\
 \hline
 $\tilde{\chi}^0_1 \tilde{\chi}^0_2 \to$ & $t\bar{t}$ &  3\% & &  & \\
                                        & $b\bar{b}$ & 23\% &   &  & \\
 \hline
 $\tilde{\chi}^0_1 \tilde{\chi}^{\pm}_1 \to$ & $t\bar{b},\bar{t}b$ & 43\% & & & \\
 \hline
 $\tilde{\chi}^0_1 \tilde{t}_1 \to$ & $th^0$ & & 1\% & 23\% & \\
                                   & $tg$ &  & 6\% & 23\%  &\\
                                   & $tZ^0$ &  &  & 5\% & \\
                                   & $bW^+$ &  & & 11\% &  \\ 
 \hline
 $\tilde{t}_1 \tilde{t}^{*}_1 \to$ & $h^0h^0$ & & 12\% & 5\%  &\\
                          		& $h^0H^0$ &  &  11\% & & \\
                              		& $Z^0A^0$ & & 7\% & & \\
                               		& $W^\pm H^\mp$ & & 13\% & & \\
                               		& $Z^0Z^0$ &  & 8\% & 2\% & \\
                               		& $W^+ W^-$ &  & 14\% & 3\% & \\
 \hline
 \multicolumn{2}{|c|}{Total} & 80\% & 72\% & 88\% & \\
 \hline
 \end{tabular}
\end{table}

In this Section, we present numerical results for three reference scenarios within the pMSSM-11,
which we introduce in Tab.\ \ref{ScenarioList}. Table \ref{ScenarioProps} shows the corresponding
relevant gaugino and squark masses, the neutralino decomposition as well as the obtained mass
of the lightest neutral (and thus SM-like) Higgs boson, the neutralino relic density computed at
the tree level with {\tt micrOMEGAs 2.4.1} \cite{Belanger:2001fz}, and the important branching ratio of
the rare $B$-meson decay $b\to s\gamma$ computed with {\tt SPheno} \cite{Porod:2003um}. Remember that
in particular $m_{\tilde{t}_2}$ is a dependent parameter in our setup and thus calculated by us
(cf.\ Sec.\ \ref{Technical}). To complete the information about the scenarios, we summarize in
Tab.\ \ref{ScenarioChannels} the most relevant (co)annihilation channels. As the physics
behind these three scenarios is quite different, we devote an individual subsection to each of
them.

Our scenarios A, B, and C are taken from the earlier publications Ref.\ \cite{Herrmann:2014kma},
Ref.\ \cite{Harz:2014gaa} and Ref.\ \cite{Harz:2014tma}, respectively, for easy comparability
and to improve on these results by additional estimates of the theoretical uncertainties.
In addition to the constraint Eq.\
(\ref{Planck}) from the relic density, we have also verified those from the Higgs mass and the
inclusive branching ratio of the decay $b\to s\gamma$. Let us note that, while in the previous
papers the trilinear coupling was entered as $T_t$, we now use the value of $A_t = T_t/Y_t$ ($Y_t$
being the top Yukawa coupling) due to a change in the \SPheno\ version. We have verified that
this change together with the replacement $T_t \to A_t$ has neither a significant impact on
the obtained mass spectrum nor on the prediction of the relic density.

\subsection{Gaugino (co)annihilation \label{Sec:GauginoCoAnni}}

\begin{figure*}
 \includegraphics[width=0.49\textwidth]{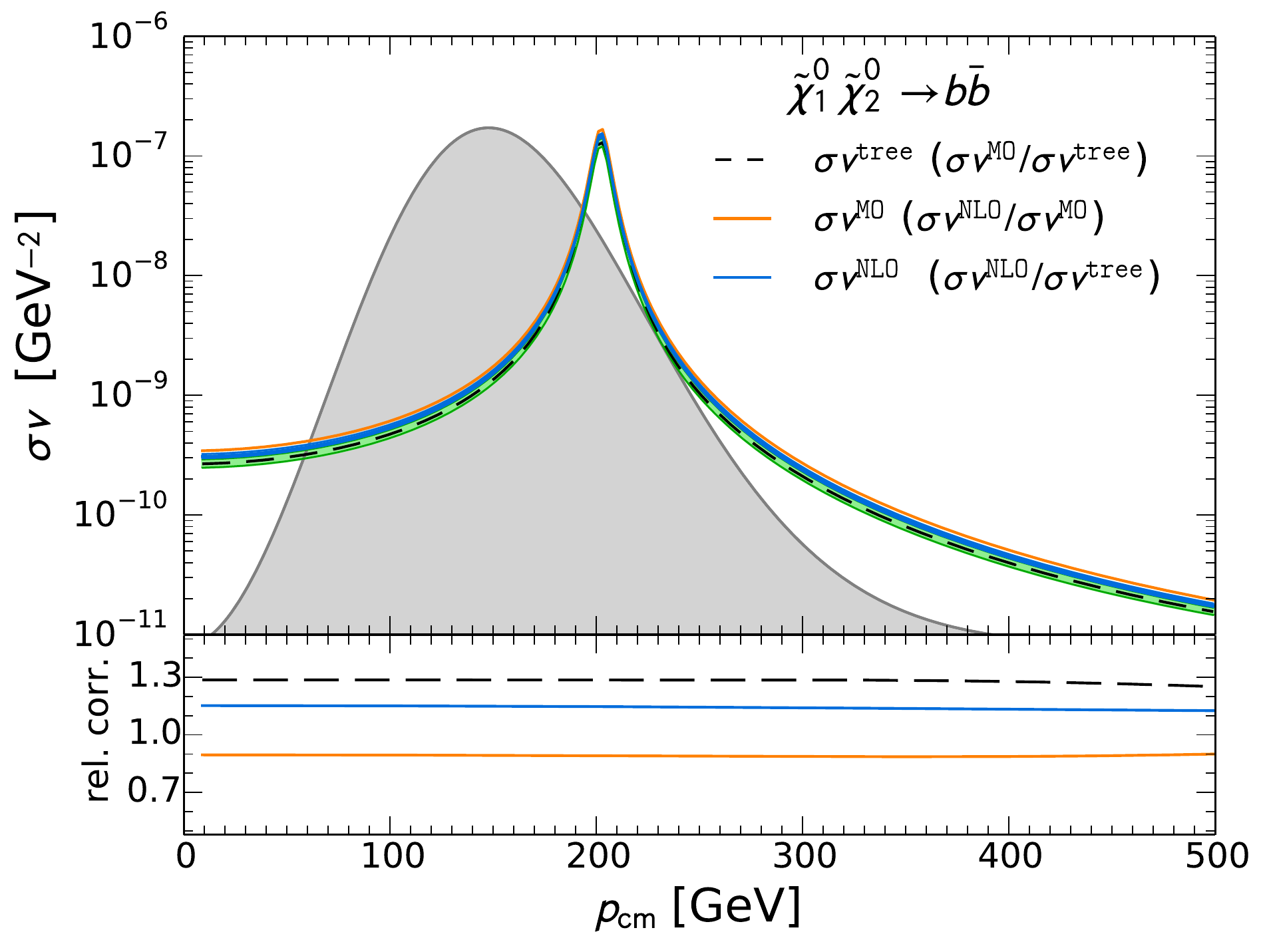}
 \includegraphics[width=0.49\textwidth]{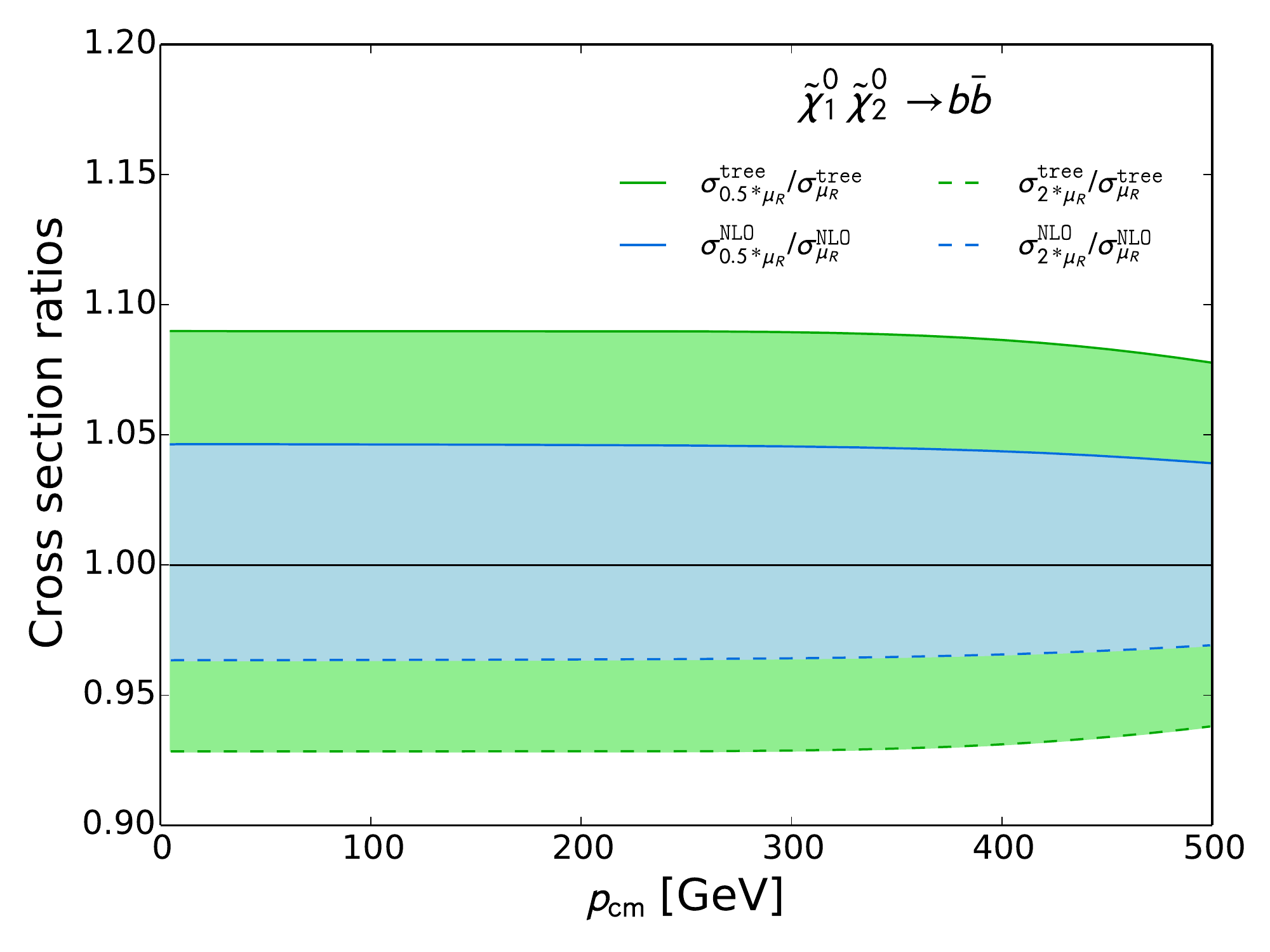}
 \includegraphics[width=0.49\textwidth]{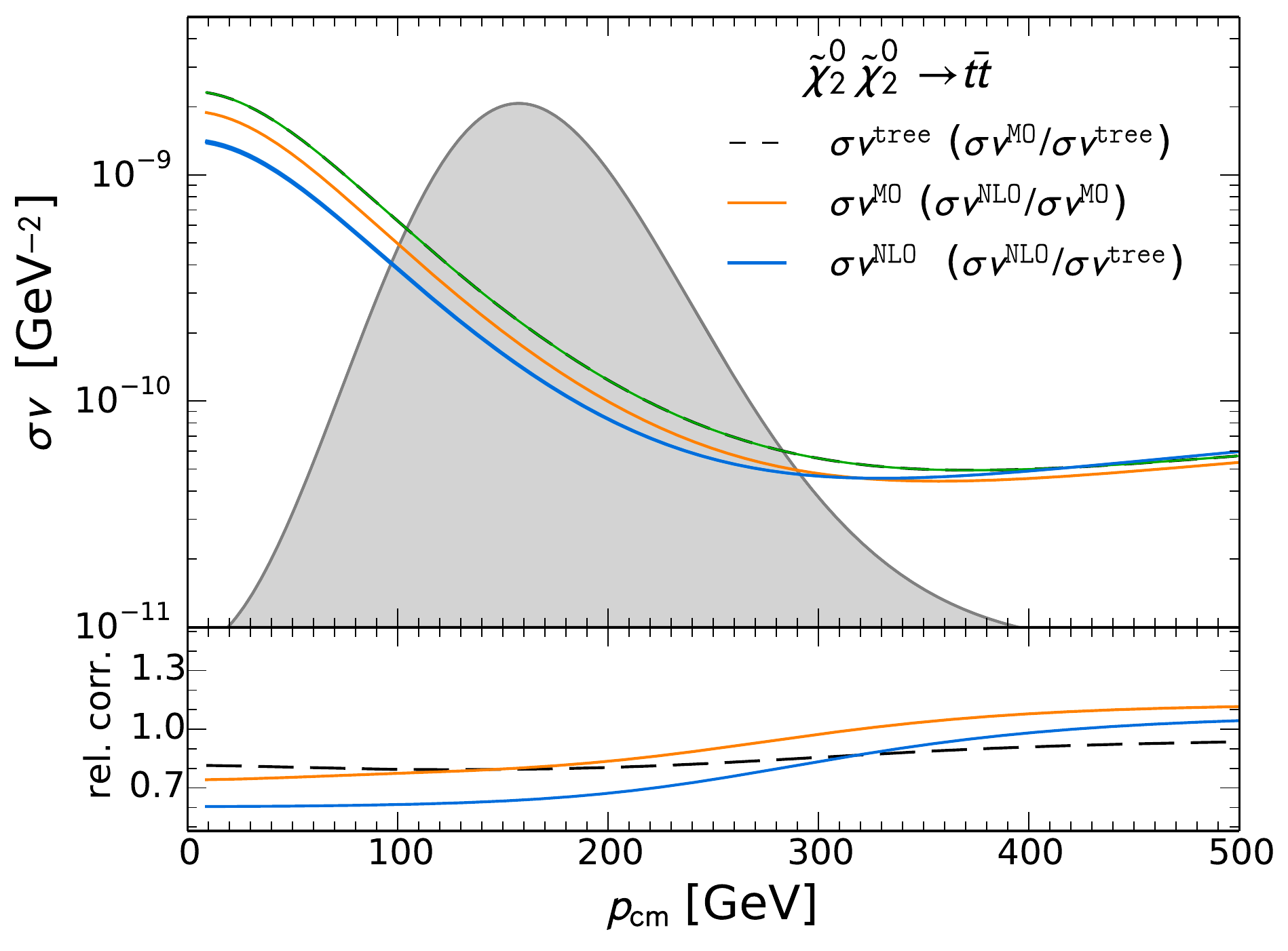}
 \includegraphics[width=0.49\textwidth]{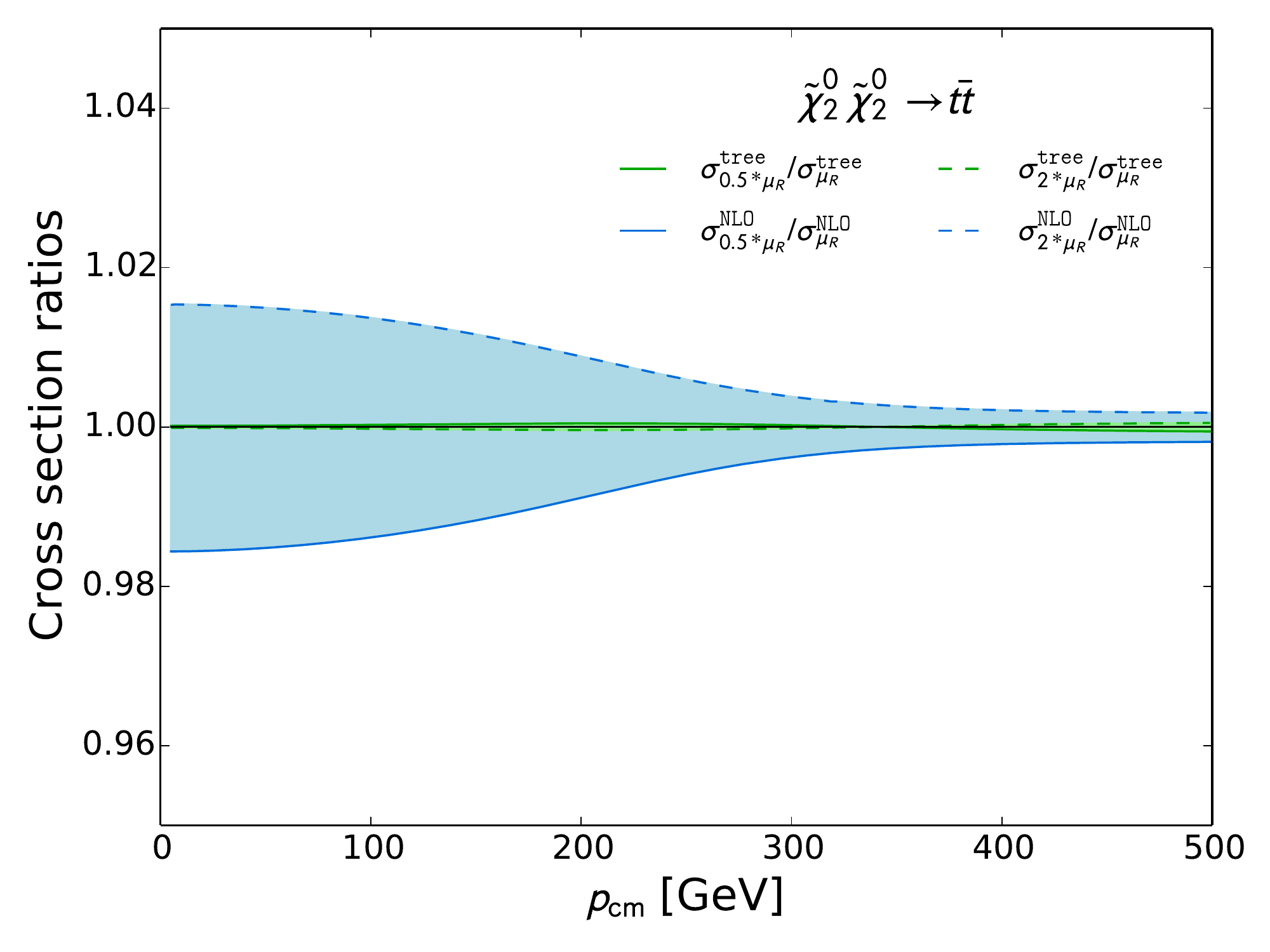}
 \caption{Cross sections of the neutralino annihilation processes $\tilde{\chi}^0_1\tilde{\chi}^0_2
 \to b\bar{b}$ (top) and $\tilde{\chi}^0_2\tilde{\chi}^0_2\to t\bar{t}$ (bottom) as a
 function of the center-of-mass momentum $p_{\rm cm}$ in scenario A. The left plots show the
 annihilation cross sections calculated at the tree level (black dashed lines) and at the
 one-loop level (blue solid lines) including the corresponding uncertainties from variations of
 the renormalization scale $\mu_R$ between $\mu_R^{\rm central}/2$ and $2\mu_R^{\rm central}$ (green
 and blue shaded bands). We also indicate the values obtained with \MO\ (orange solid lines).
 The right plots show the cross sections normalized to their values obtained with the central
 renormalization scale $\mu_R^{\rm central}=1$ TeV.}
 \label{fig:ScaleDependenciesScenI}
\end{figure*}

\begin{figure}
 \includegraphics[width=0.49\textwidth]{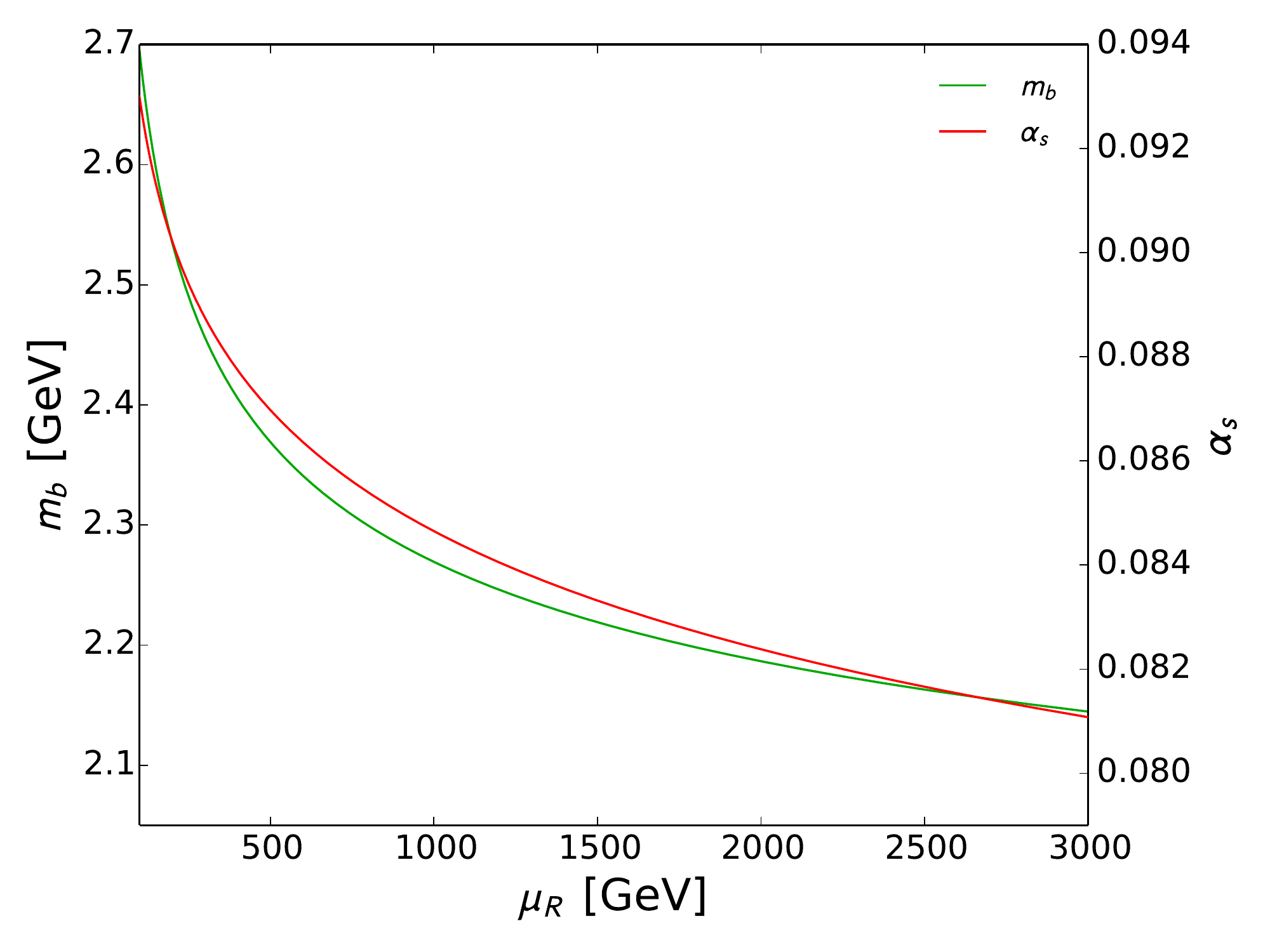}
 \caption{Dependence of the bottom-quark mass $m_b$ (green line, left ordinate) and of the strong
 coupling constant $\alpha_s$ (red line, right ordinate) on the renormalization scale $\mu_R$ in
 the MSSM $\overline{\rm DR}$ scheme in scenario A.}
 \label{fig:ScaleDependence}
\end{figure}

We begin our numerical investigation by studying the cross section of the process
$\tilde{\chi}_1^0\tilde{\chi}_2^0 \rightarrow b\bar{b}$ in scenario A. We do not reproduce here
a detailed phenomenological discussion of this scenario, as it can be found in Ref.\
\cite{Herrmann:2014kma}. The upper left graph of Fig.\ \ref{fig:ScaleDependenciesScenI} presents the
cross section times the relative neutralino velocity $v$ for the mentioned process calculated
with \CHep-based \cite{Pukhov:2004ca} \MO\ \cite{Belanger:2001fz} (orange solid line), \DMNLO\
at tree level (black dashed line), and NLO (blue solid line) at $\mu_R^{\rm central} = 1$ TeV as
a function of the center-of-mass momentum $p_{\mathrm{cm}}$. Note that in our code the
widths of unstable particles are always active, whereas in \MO\ they are switched on only
in a rather narrow interval around the resonance. In order to compare our calculation with the
one implemented in \MO, we have modified the treatment of the width in \CHep\ such that it is
taken into account over the full range of $p_{\mathrm{cm}}$. For the calculation of the relic
density, however, we have not modified the treatment of the width in \CHep.
The general shape of the curves is dominated by a large $A^0$ resonance at $p_{\mathrm{cm}} \approx
200$ GeV in the vicinity of the maximum of the thermal dark matter velocity distribution (gray
shaded contour), which makes this process relevant for neutralino relic density calculations
(see Tab.\ \ref{ScenarioChannels}). The difference between the two tree-level calculations can be
traced back to the fact that \MO\ uses effective masses and couplings, while we do not. The lower
subplot shows the ratios of the three cross sections (cf.\ the second items in the legend).
The new features in this plot with respect to the analysis presented in Ref.\ \cite{Herrmann:2014kma}
are the green and blue shaded bands associated with the dashed black and blue solid lines. These
bands are limited by calculations with renormalization scales
of $\mu_R = 0.5$ TeV and $\mu_R = 2$ TeV, respectively, so that the area between them indicates
the theoretical uncertainty due to renormalization scale variations. Remember that the input
scale used for the pMSSM-11 parameters $\tilde{M}\equiv\mu_R^{\rm central} = 1$ TeV remains
unchanged, as we do not want to change the underlying scenario (see Sec.\ \ref{Technical}).
We observe a rather large scale dependence of our tree-level cross section (green shaded band)
and a smaller dependence at NLO (blue shaded band). The bands overlap, as is visible at around
$p_{\mathrm{cm}} = 100$ GeV, but the central NLO result does not fall in the tree-level band.
In other words, the impact of NLO radiative corrections on the final result is larger than the
uncertainty estimated from the LO scale dependence. This is typical for a process that is of
purely electroweak origin at LO. The orange \MO\ results lies outside of both
bands. This means that our full NLO cross section differs from the effective tree-level result
obtained by \MO\ even after including the scale uncertainty.

To enhance the visibility of the scale uncertainties, in particular near the resonance, we show
in the upper right plot of Fig.\ \ref{fig:ScaleDependenciesScenI} ratios of the tree-level and NLO
cross sections at $\mu_R=(0.5;2)\times\mu_R^{\rm central}$ over the central results. We find an
increase of $\sim 9\%$ of the tree-level cross section when changing from $\mu_R = 1$ TeV to
$\mu_R = 0.5$ TeV and a decrease of $\sim 7\%$ when working with $\mu_R = 2$ TeV (green shaded
band). The shifts are almost constant and decrease only slightly at high $p_{\mathrm{cm}}$. The
reason is that the dominant subprocess here is the $s$-channel process $\tilde{\chi}_1^0
\tilde{\chi}_2^0\rightarrow A^0\rightarrow b\bar{b}$, as we now explain: The coupling
$A^0b\bar{b}$ is of Yukawa type and proportional to $m_b$. Therefore, the total subprocess is
proportional to $m_b^2$. As the bottom mass is handled as a \DRbar\ parameter in our code (cf.\
Sec.\ \ref{Technical}), its scale dependence directly translates to the cross section in this
case.

This scale dependence is illustrated in Fig.\ \ref{fig:ScaleDependence}. More precisely, the
ratios $(\frac{m_b(\mu_R =2\,\mathrm{TeV})}{m_b(\mu_R = 1\,\mathrm{TeV})})^2 \approx 0.93$ and
$(\frac{m_b(\mu_R = 0.5\,\mathrm{TeV})}{m_b(\mu_R = 1\,\mathrm{TeV})})^2 \approx 1.09$ reflect the
observed shifts very well. The slight decrease of these shifts at high $p_{\mathrm{cm}}$ in the
upper right plot of Fig.\ \ref{fig:ScaleDependenciesScenI} has a related origin. Since the
relative contributions of different subprocesses depend on $p_{\mathrm{cm}}$ (see Tab.\ IV in Ref.\
\cite{Herrmann:2014kma}), the subprocess $\tilde{\chi}_1^0\tilde{\chi}_2^0\rightarrow
A^0\rightarrow b\bar{b}$ becomes less dominant at high $p_{\mathrm{cm}}$, and other subprocesses,
which are not proportional to $m_b$, start to contribute as well. Hence the influence of
$m_b$ and its scale dependence on the whole process decreases.
The scale dependence is reduced at NLO, as shown by the blue shaded band in the upper right plot
of Fig.\ \ref{fig:ScaleDependenciesScenI}. This is exactly as expected: Including virtual
corrections, in particular vertex corrections to the $A^0b\bar{b}$ Yukawa coupling, reduces the
scale depence of the resulting cross section. The remaining uncertainty amounts to less than five
percent.

We continue with the lower part of Fig.\ \ref{fig:ScaleDependenciesScenI}. The lower left plot
shows the cross section of the non-resonant process $\tilde{\chi}_2^0\tilde{\chi}_2^0\rightarrow
t\bar{t}$. In this case, the Yukawa couplings contain the top-quark mass, which is defined as a
pole mass in our code and therefore scale independent. Due to its rather small cross section,
this process is irrelevant in terms of relic density calculations and hence not listed in Tab.\
\ref{ScenarioChannels}. Nevertheless it proves useful for illustrating the scale dependence of
our code.

As before, the black dashed and blue solid lines are associated with the green and blue shaded
bands corresponding to the tree-level and NLO cross sections, respectively, at $\mu_R = 0.5$ TeV
and $\mu_R = 2$ TeV. However, these bands are almost invisible in this case, as they
overlap with the original lines. This fact indicates a much smaller scale dependence of the
process than previously, which is confirmed in the lower right plot of Fig.\
\ref{fig:ScaleDependenciesScenI}. The (purely electroweak) tree-level cross section is basically
scale independent, i.e.\ the green shaded band coincides with a nearly constant value of one. 
At NLO (blue shaded band), the scale dependence is now increased, to up to two percent at low
$p_{\mathrm{cm}}$. This is due to the scale dependence of $\alpha_s$ introduced by our SUSY QCD
corrections and depicted in Fig.\ \ref{fig:ScaleDependence}. Although this seems to worsen the
reliability of the calculation at first sight, the contrary is true: only at NLO it becomes
possible to quantify the theoretical error for the first time.
In this channel, the NLO corrections tend to decrease the cross section, i.e.\ the blue solid
line lies below the black dashed line and the NLO/LO ratio below one. This effect is reduced at
higher $p_{\mathrm{cm}}$, so that the NLO scale dependence decreases significantly.
Note that an explicit dependence on $\alpha_s$ is also introduced in the process $\tilde{\chi}_1^0
\tilde{\chi}_2^0\rightarrow b\bar{b}$ discussed at the beginning of this Section. There,
however, it is completely overshadowed by the dominant scale dependence of $m_b$. We will
encounter similar competing scale dependencies (e.g.\ on $\alpha_s$ and $A_t$) in other
processes later.

\begin{figure}
 \includegraphics[width=0.49\textwidth]{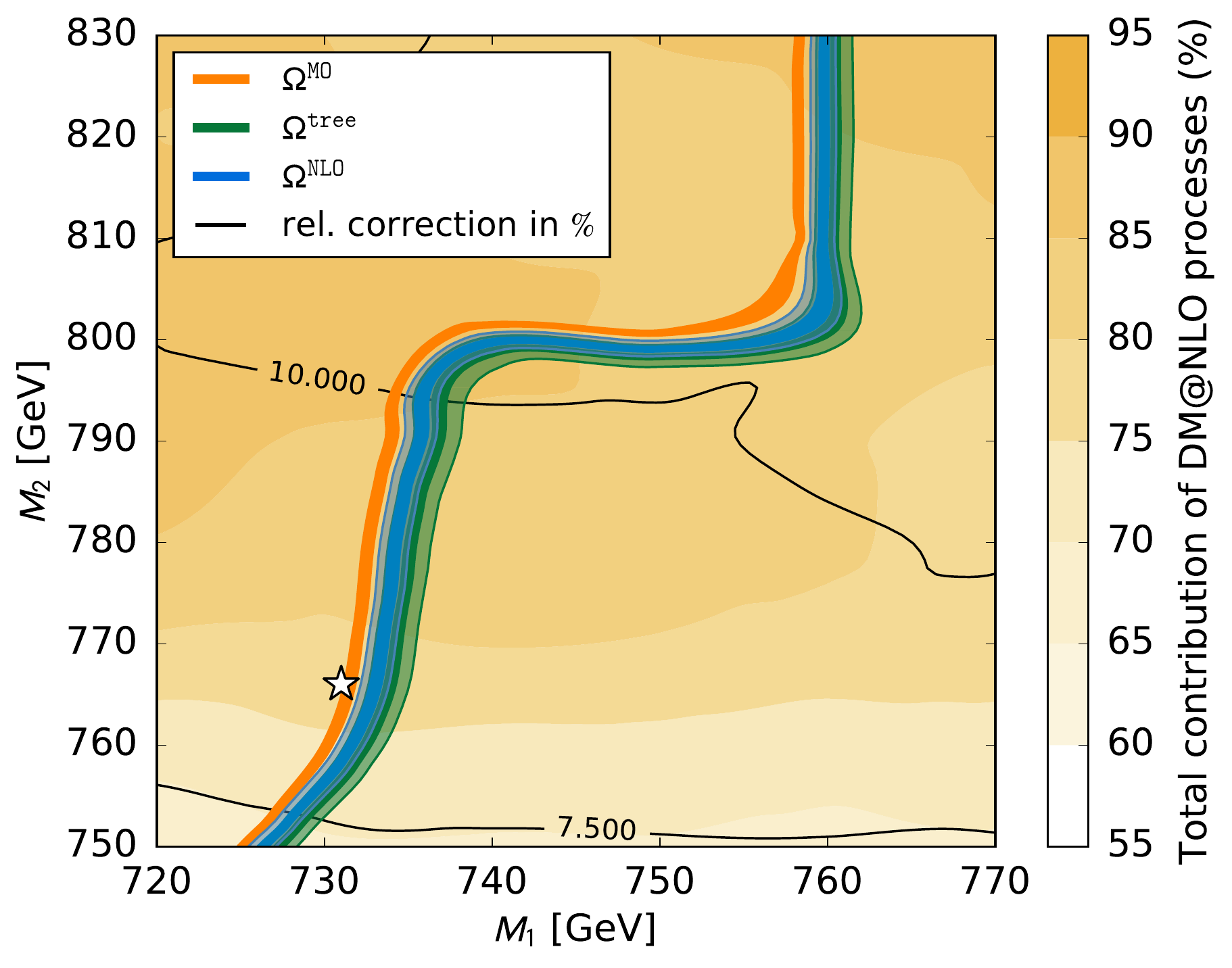}
 \caption{Cosmologically preferred regions with respect to Eq.\ \eqref{Planck} in the
 $M_1$-$M_2$-plane surrounding scenario A (white star). The solid bands correspond to the limits
 of Eq.\ \eqref{Planck}, while the shaded bands include an additional variation of the
 renormalization scale $\mu_R$ between $\mu_R^{\rm central}/2$ and $2\mu_R^{\rm central}$. The shades of
 yellow indicate the percentages of the total annihilation cross section which
 have been corrected in our calculation, while the solid lines represent contours of the relative
 correction to the relic density in percent.}
 \label{fig:RelicScenA}
\end{figure}

We close this subsection with a discussion of the scale dependence of the neutralino relic
density. In Fig.\ \ref{fig:RelicScenA} we show the relic density in the $M_1$--$M_2$-plane
surrounding scenario A. The three colored, solid lines represent the part of the parameter space
which leads to a neutralino relic density compatible with the Planck limits given in Eq.\
\eqref{Planck}. These lines are rather thin, which reflects the high precision of the Planck
measurement. For the orange line we used the standard \MO\ routine, the green one corresponds to
our tree-level calculation, and the blue one represents our full one-loop calculation. The last
two lines are surrounded by green and blue shaded bands, which correspond to changing the
renormalization scale to 0.5 or 2 TeV. The shades of yellow denote the relative fraction of
processes we correct with \DMNLO\ and the black lines the relative correction to the relic
density.
We observe that the tree-level and NLO results clearly overlap within the scale uncertainty.
This is not unexpected, as the relative shift from \MO\ (and similarly LO) to NLO 
in this scenario happens to be rather small
($5-10$\%). Furthermore note that the scale dependence of the relic density reduces at NLO,
i.e.\ the blue shaded band surrounding the blue line is smaller than the green
band surrounding the green line. This can be understood as follows:
The processes we correct with \DMNLO\ account for 80\% in this part of the parameter space. The
most important ones are $\tilde{\chi}_1^0\tilde{\chi}_2^0\rightarrow b\bar{b}$, $\tilde{\chi}_1^0
\tilde{\chi}_1^0\rightarrow b\bar{b}$ and $\tilde{\chi}_1^+\tilde{\chi}_1^0\rightarrow t\bar{b}$
(cf.\ Tab.\ \ref{ScenarioChannels} and Ref.\ \cite{Herrmann:2014kma}) and mainly take place via
$s$-channel Higgs exchanges. The tree-level couplings entering these processes all contain
Yukawa couplings proportional to the bottom quark $m_b$. Therefore, the discussion of the cross
section of the process $\tilde{\chi}_1^0\tilde{\chi}_2^0\rightarrow b\bar{b}$ above translates to
the relic density, and the dominant source of scale dependence is again the \DRbar\ bottom mass.
This scale dependence is decreased at NLO, when vertex and other NLO corrections are included.
Our NLO result (blue solid line) differs from the \MO\ result (orange solid curve) even after
including the scale uncertainty (blue shaded band), which we have already previously traced back
to different treatments of the third-generation quark masses \cite{Herrmann:2014kma}.

An extraction of the pMSSM-11 parameters from the Planck relic density with \MO\ would lead to
$M_1=731\pm1$ GeV and $M_2=766\pm1$ GeV and a precision of about 1 per mille, while our
calculations would rather imply $M_1=731\pm2$ GeV and $M_2=759\pm2$ GeV, i.e.\ a shift in
$M_2$ by 7 GeV or 1\% with an uncertainty of two per mille, two times bigger
than naively believed.\footnote{To obtained the total error, the experimental and theoretical
uncertainties would, of course, have to be added in quadrature.}

\subsection{Stop-antistop annihilation}

\label{QQChapter}

\begin{figure*}
 \includegraphics[width=0.49\textwidth]{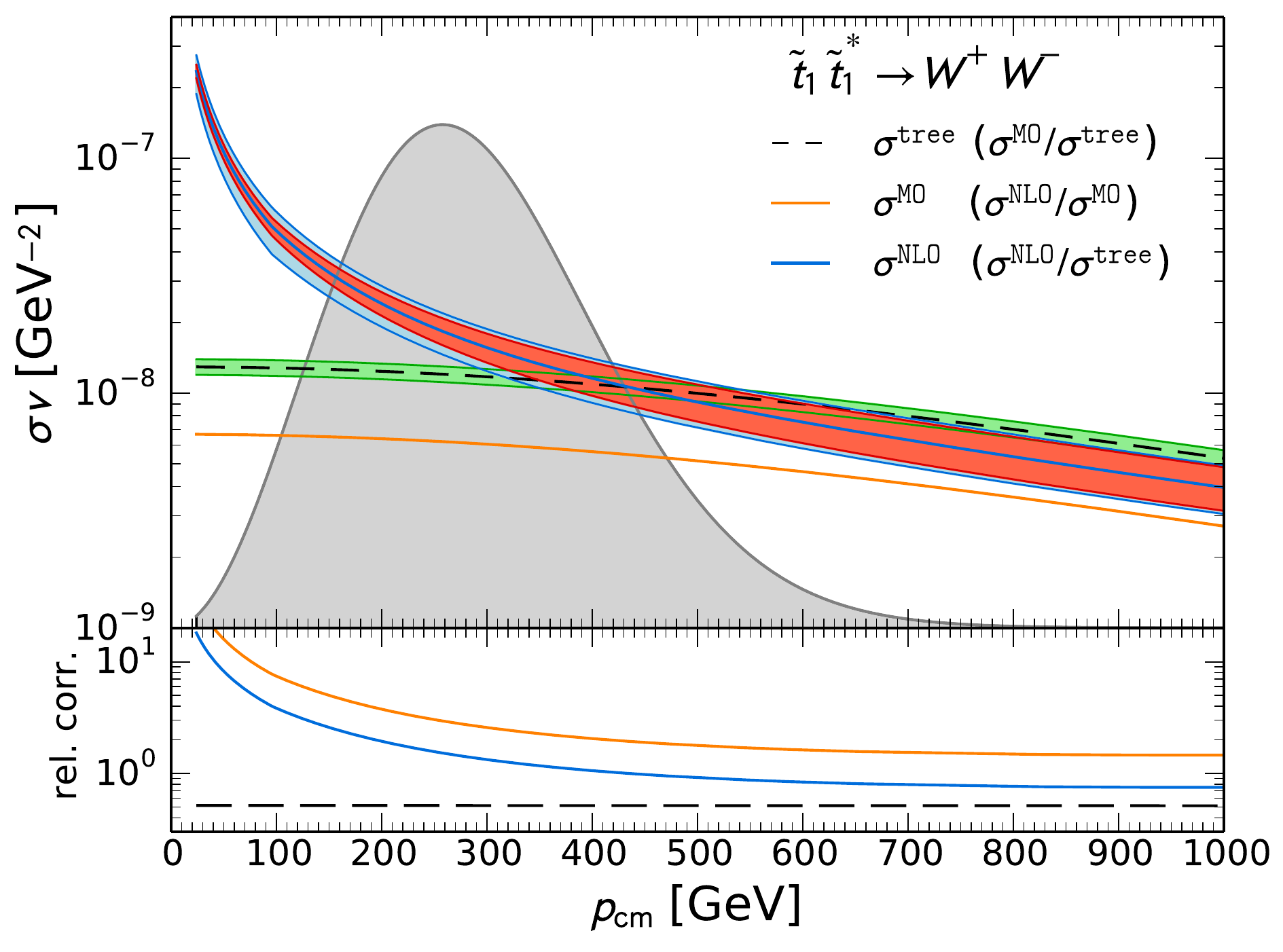}
 \includegraphics[width=0.49\textwidth]{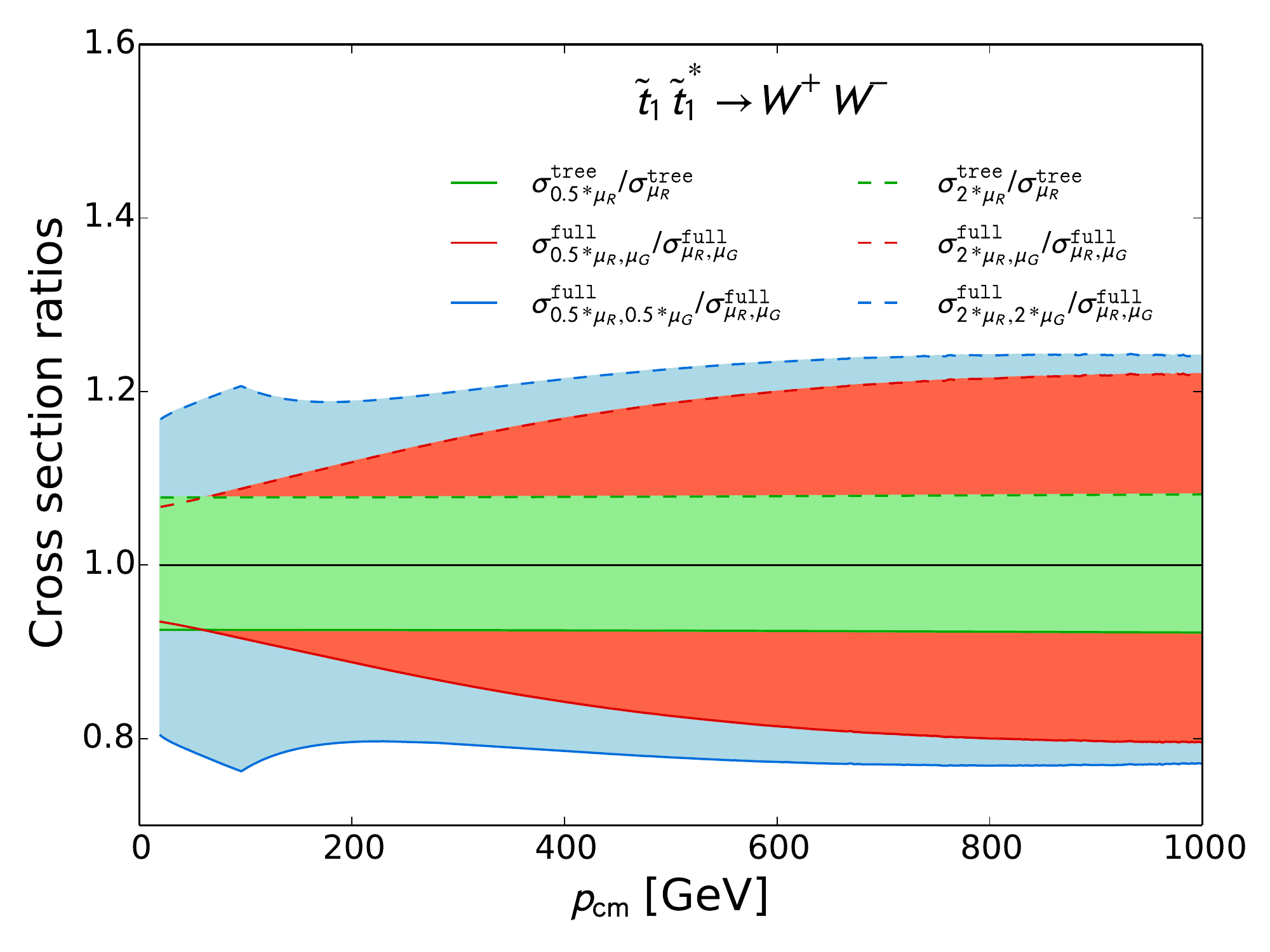}
 \caption{Cross sections of the stop-antistop annihilation process $\tilde{t}_1\tilde{t}_1^*\to
 W^+W^-$ as a function of the center-of-mass momentum $p_{\rm cm}$ in scenario B. The left plot
 shows the annihilation cross sections calculated at the tree level (black dashed line) and at
 the one-loop level with Coulomb resummation (blue solid line) including the corresponding
 uncertainties from simultaneous variations of the renormalization scale $\mu_R$ and the Coulomb
 scale $\mu_C$ by factors of two around the central scales (green and blue shaded bands). Also
 shown is the uncertainty from variations of $\mu_R$ alone (red shaded band). We also
 indicate the value obtained with \MO\ (orange solid line). The right plot shows the cross
 sections normalized to their values obtained with central renormalization and Coulomb scales
 $\mu_R^{\rm central}$ and $\mu_C^{\rm central}$.}
 \label{fig:ScaleDependenciesScenB}
\end{figure*}

We continue our analysis with scenario B, where stop-antistop annihilation is dominant (see
Tab.\ \ref{ScenarioChannels}). This scenario has been introduced and studied as scenario II in
Ref.\ \cite{Harz:2014gaa}. As an example, we investigate in the left plot of Fig.\
\ref{fig:ScaleDependenciesScenB} the cross section of the process $\tilde{t}\tilde{t}^*
\rightarrow W^+W^-$. The color codes for the \MO\ (orange solid lines) and our central tree-level
cross sections (black dashed line) are as before. Our full result (blue solid line) includes now
NLO and Coulomb corrections, calculated at a renormalization scale of $\mu_R^{\rm central} = 1$ TeV
and a Coulomb scale of $\mu_C^{\rm central}=\max\{\mu_B,2m_{\tilde{t}_1}v\} $ (cf.\ Sec.\
\ref{Technical}).

The Coulomb corrections lead to a strong enhancement of the cross section of up to a factor of
ten in particular at small $p_{\mathrm{cm}}$, as observed also in Figs.\ 6 and 8 of Ref.\
\cite{Harz:2014gaa}.
The Coulomb enhancement region overlaps significantly with the Boltzmann velocity distribution
and thus contributes in an important way to the dark matter relic density.
At large $p_{\rm cm}$, our full cross section falls below the one at tree level, leading to a
$K$-factor of about 0.75 (cf.\ the left lower subplot in Fig.\ \ref{fig:ScaleDependenciesScenB}),
and approaches the one in \MO, but still differs from it due to different definitions of the
top-quark mass. Their effect is, however, reversed and reduced by the one-loop corrections
included in our full calculation.

\begin{figure}
 \includegraphics[width=0.49\textwidth]{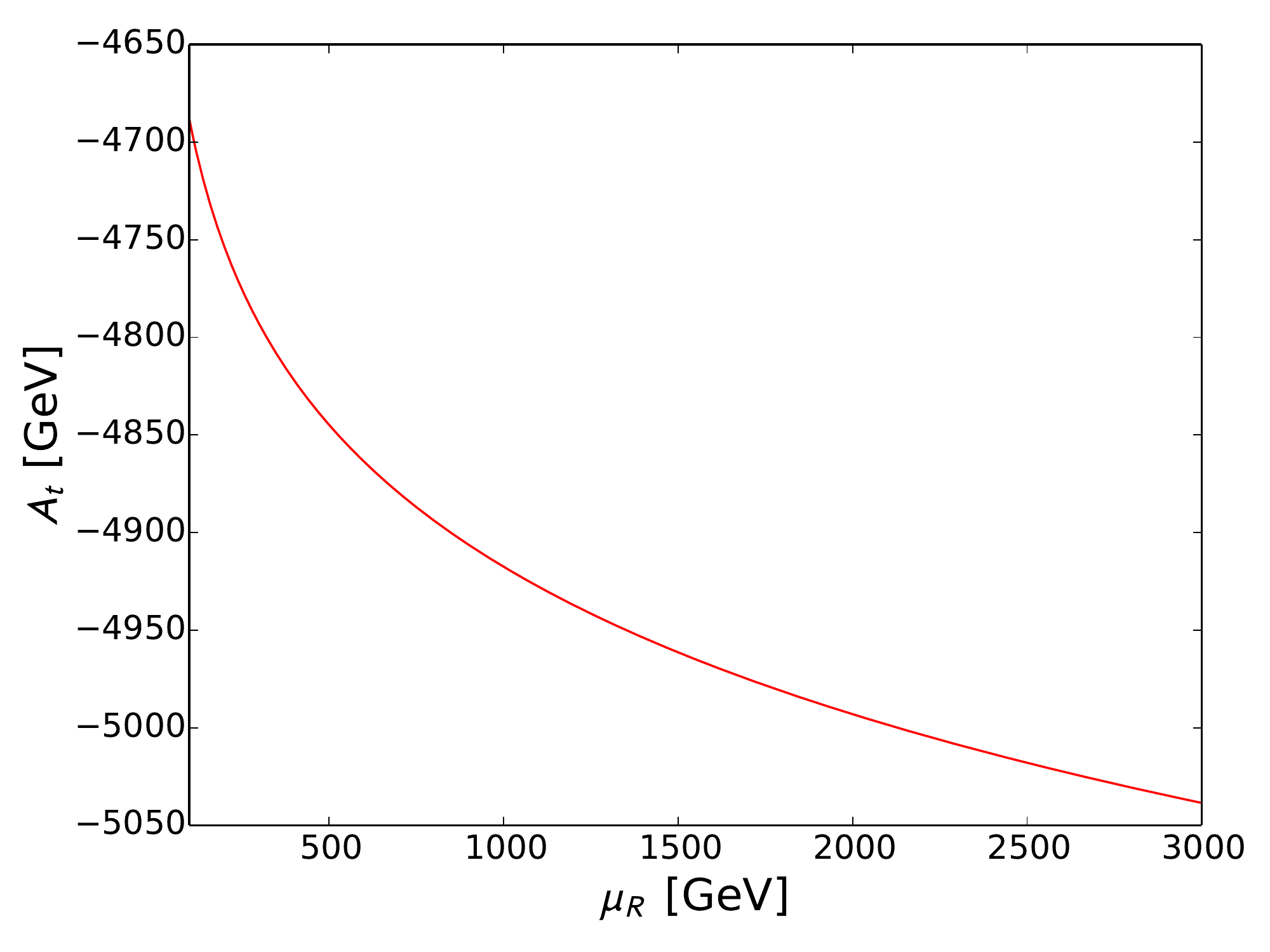}
 \caption{Dependence of the trilinear coupling $A_t$ (red line) on the renormalization scale
 $\mu_R$ in scenario B.}
 \label{fig:ScaleDependenceAt}
\end{figure}

As before, the green shaded band indicates the theoretical uncertainty of the tree-level result
induced by variations of $\mu_R$ by factors of two. Here, the renormalization scale influences the
tree-level cross section through the trilinear coupling $A_t$, on which the squark mixing matrices
and thus the squark-squark-vector couplings of the dominant $t$- and $u$-channel squark-exchange
subprocesses depend (cf. table IV in Ref.\ \cite{Harz:2014gaa}). If we keep $A_t$ fixed, the tree-level
result becomes scale independent.
We show the scale dependence of $A_t$ explicitly in Fig.\ \ref{fig:ScaleDependenceAt}.

\begin{figure}
 \includegraphics[width=0.49\textwidth]{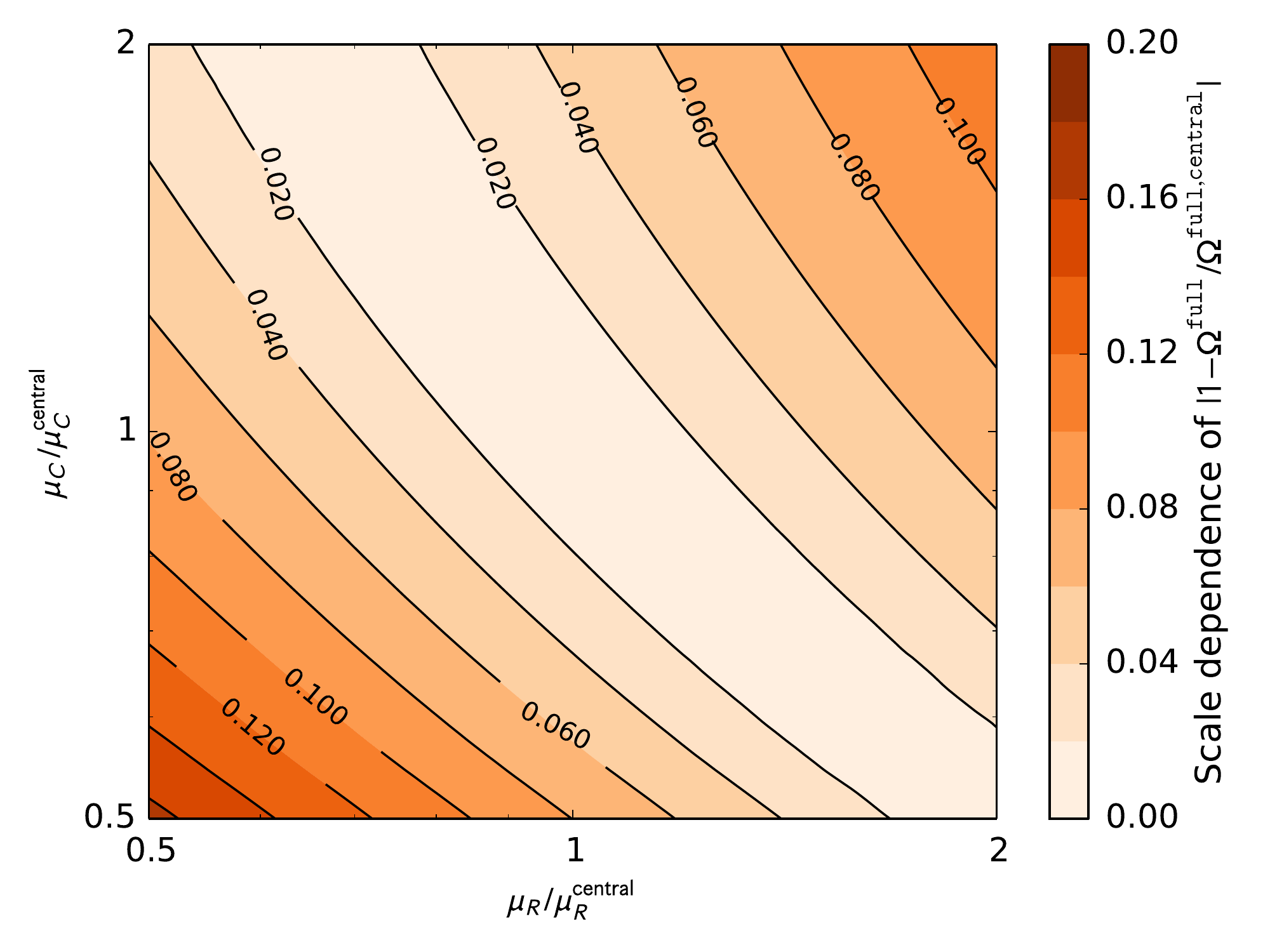}
 \caption{Dependence of the relic density on the scales $\mu_R$ and $\mu_C$ in scenario B.}
\label{fig:Temperaturplot}
\end{figure}

In contrast, the blue shaded band in Fig.\ \ref{fig:ScaleDependenciesScenB} now indicates
the simultaneous variations of $\mu_R$ and $\mu_C$ by factors of two, i.e.\ from $\mu_R=\mu_C=
0.5$ TeV to 2 TeV. This is the most conservative procedure to combine the two uncertainties,
as can be seen from the dependence of the relic density on the two scales in Fig.\
\ref{fig:Temperaturplot}.

The dependence of our full cross section on $\mu_R$ alone is shown in Fig.\
\ref{fig:ScaleDependenciesScenB} as a red shaded band. As one observes from the right-hand
plot, it dominates the full theoretical uncertainty with an error of about $\pm20\%$ at large
$p_{\rm cm}$. The one-loop corrections there depend implicitly on $\mu_R$ through the newly
introduced strong coupling constant $\alpha_s$ and explicitly (logarithmically) through the NLO
corrections. In contrast to scenario A, these dependencies are sizeable in scenario
B and not completely overshadowed by the leading-order dependence on $m_b$ discussed there.
The uncertainty induced by $\mu_R$ on the full cross section falls to the level of about
$\pm7\%$ at low $p_{\rm cm}$, where it becomes comparable to the constant tree-level uncertainty.
However, there the Coulomb corrections and their theoretical uncertainty, induced by variations of
$\mu_C$, become important as expected, and they increase the full higher-order theoretical
uncertainty again to about $\pm20\%$.

\begin{figure}
 \includegraphics[width=0.49\textwidth]{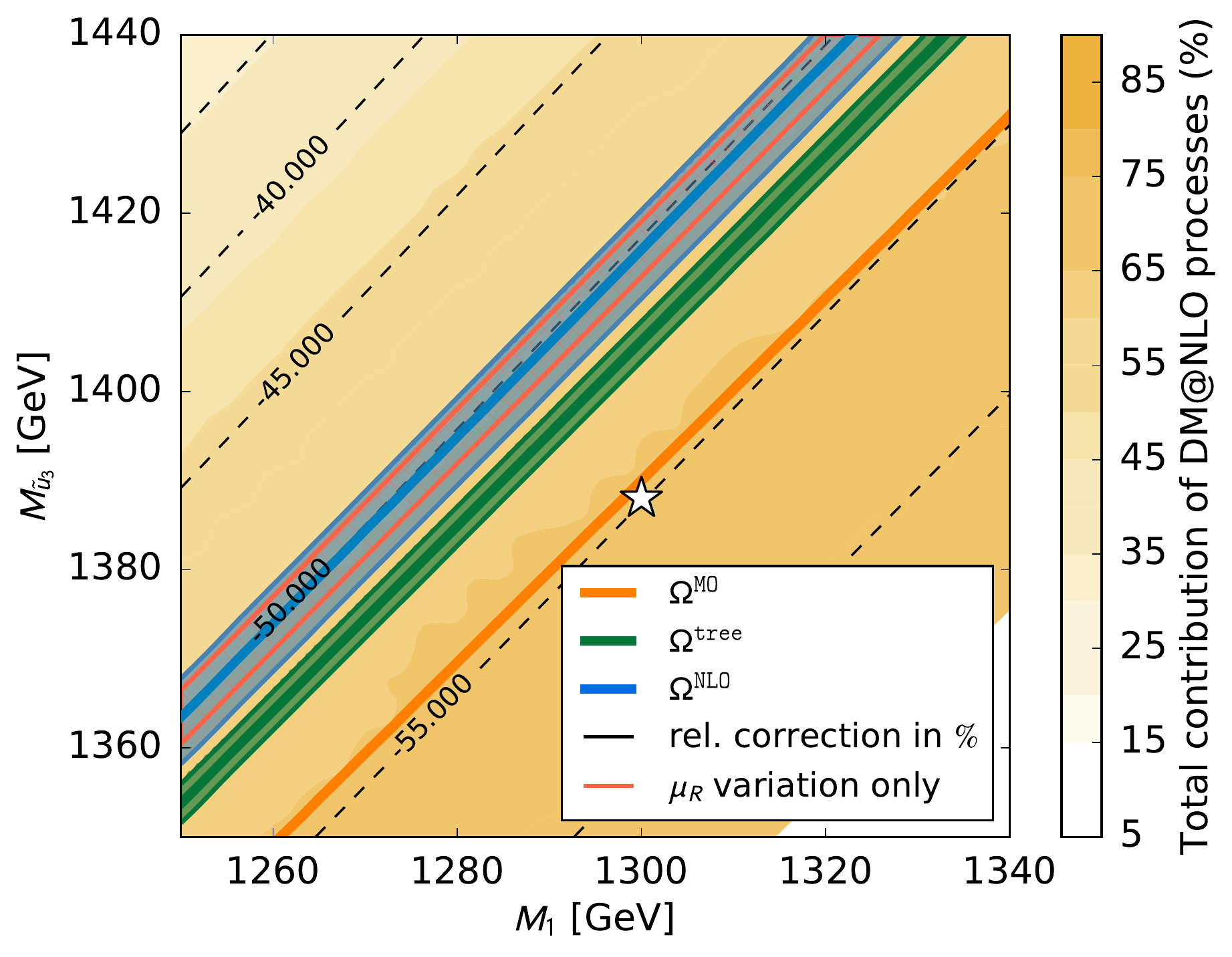}
 \caption{Cosmologically preferred regions with respect to Eq.\ \eqref{Planck} in the
 $M_1$-$M_{\tilde{u}_3}$-plane surrounding scenario B (white star). The solid bands correspond to
 the limits of Eq.\ \eqref{Planck}, while the shaded bands include additional
 variations of the renormalization scale $\mu_R$ and the Coulomb scale $\mu_C$ by factors of
 two around the central scales. The shades of
 yellow indicate the percentages of the total annihilation cross section which
 have been corrected in our calculation, while the dashed lines represent contours of the relative
 correction to the relic density in percent.}
\label{fig:relicScenB}
\end{figure}

To conclude this subsection, we show in Fig.\ \ref{fig:relicScenB} a relic-density scan in the
$M_1-M_{\tilde{u}_3}$ plane around our representative scenario B (white star). In the whole region,
our calculations correct more than 60\% of the contributing subprocesses (shades of yellow). The
tree-level result  (green line) differs again visibly from the \MO\ result (orange
line) defined in a different way and exhibits only a small renormalization scale dependence
(green shaded band) induced by the trilinear coupling $A_t$. The combined scale uncertainty from
variations of $\mu_R$ and $\mu_C$ of the full result (blue shaded band) is larger and
significantly broadens the 1$\sigma$-band representing the experimental error from Planck.
Due to the important Coulomb enhancement, the higher-order uncertainty band does not overlap
with the one at tree level, but since these corrections are resummed to all orders, this does not
imply that they are unreliable.

An extraction of the pMSSM-11 parameters from the Planck relic
density with \MO\ would lead to $M_1=1300\pm1$ GeV and $M_{\tilde{u}_3}=1388\pm1$ GeV and a precision
of about 1 per mille, while our calculations would rather imply $M_1=1300\pm5$ GeV and
$M_{\tilde{u}_3}=1415\pm5$ GeV, i.e.\ a shift in $M_{\tilde{u}_3}$ (or equivalently $M_1$) by 30 GeV
or 2\% with an uncertainty of $0.5\%$, five times bigger than naively believed.

\subsection{Neutralino-stop coannihilation}

\begin{figure*}
 \includegraphics[width=0.49\textwidth]{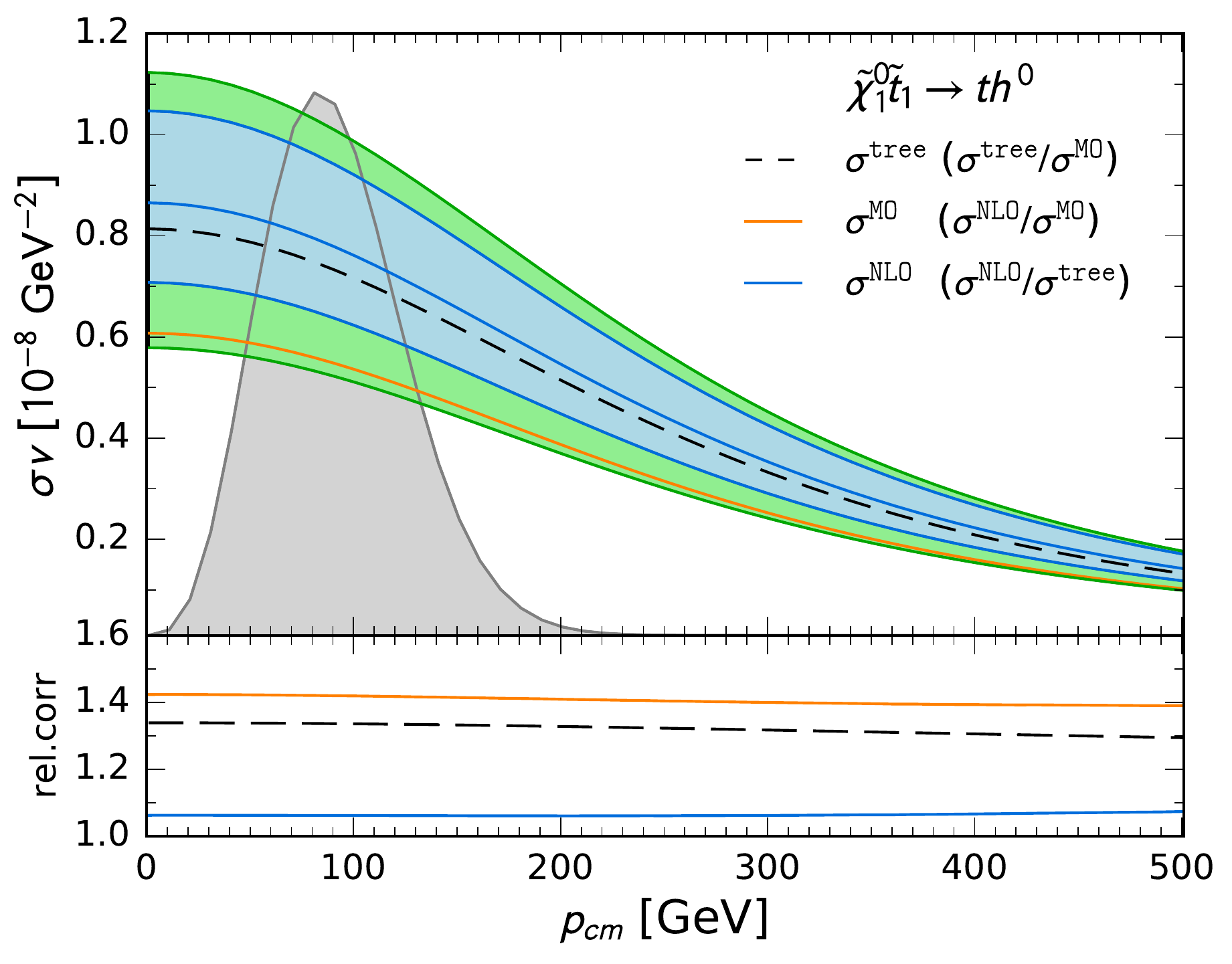}
 \includegraphics[width=0.49\textwidth]{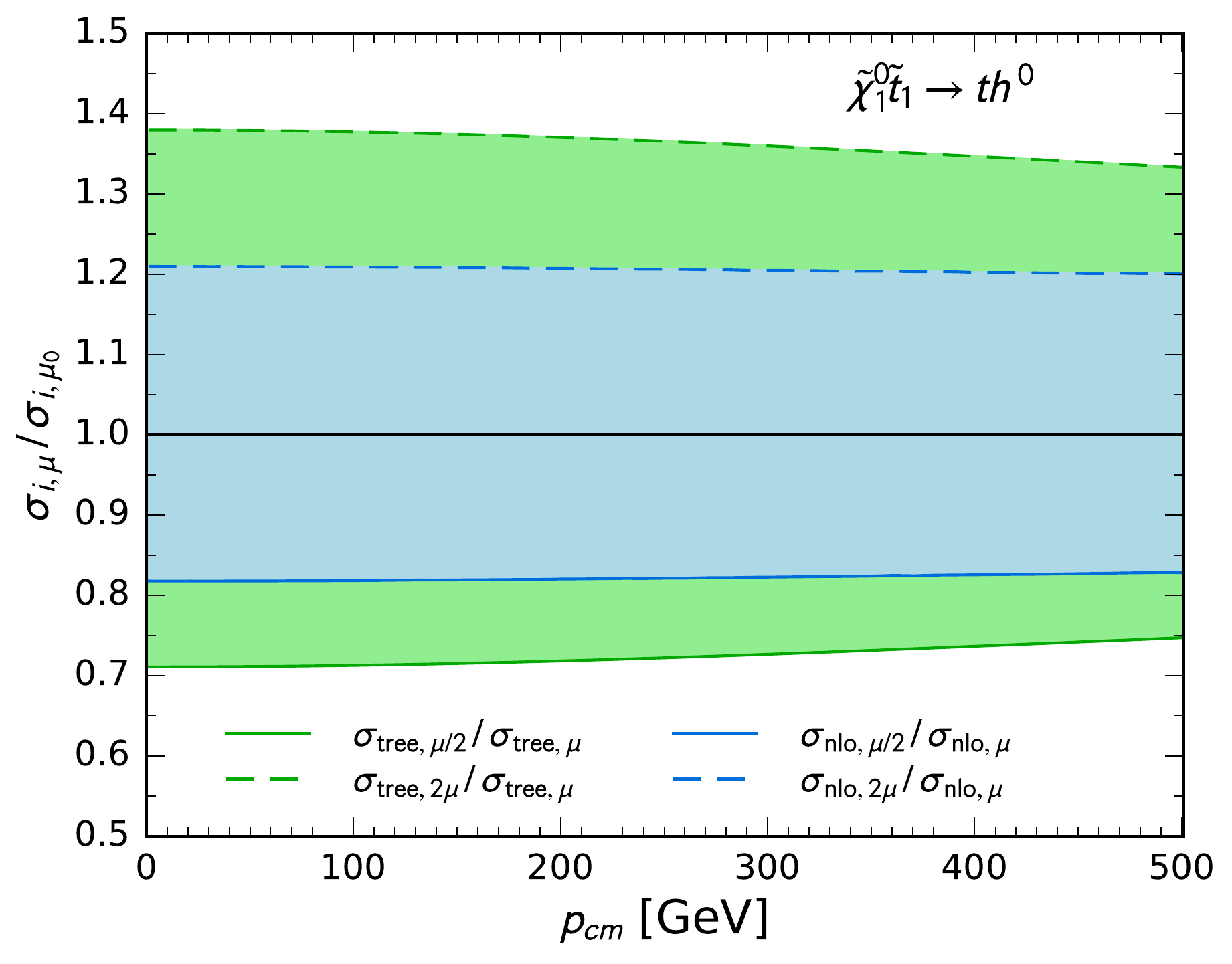}\\
 \includegraphics[width=0.49\textwidth]{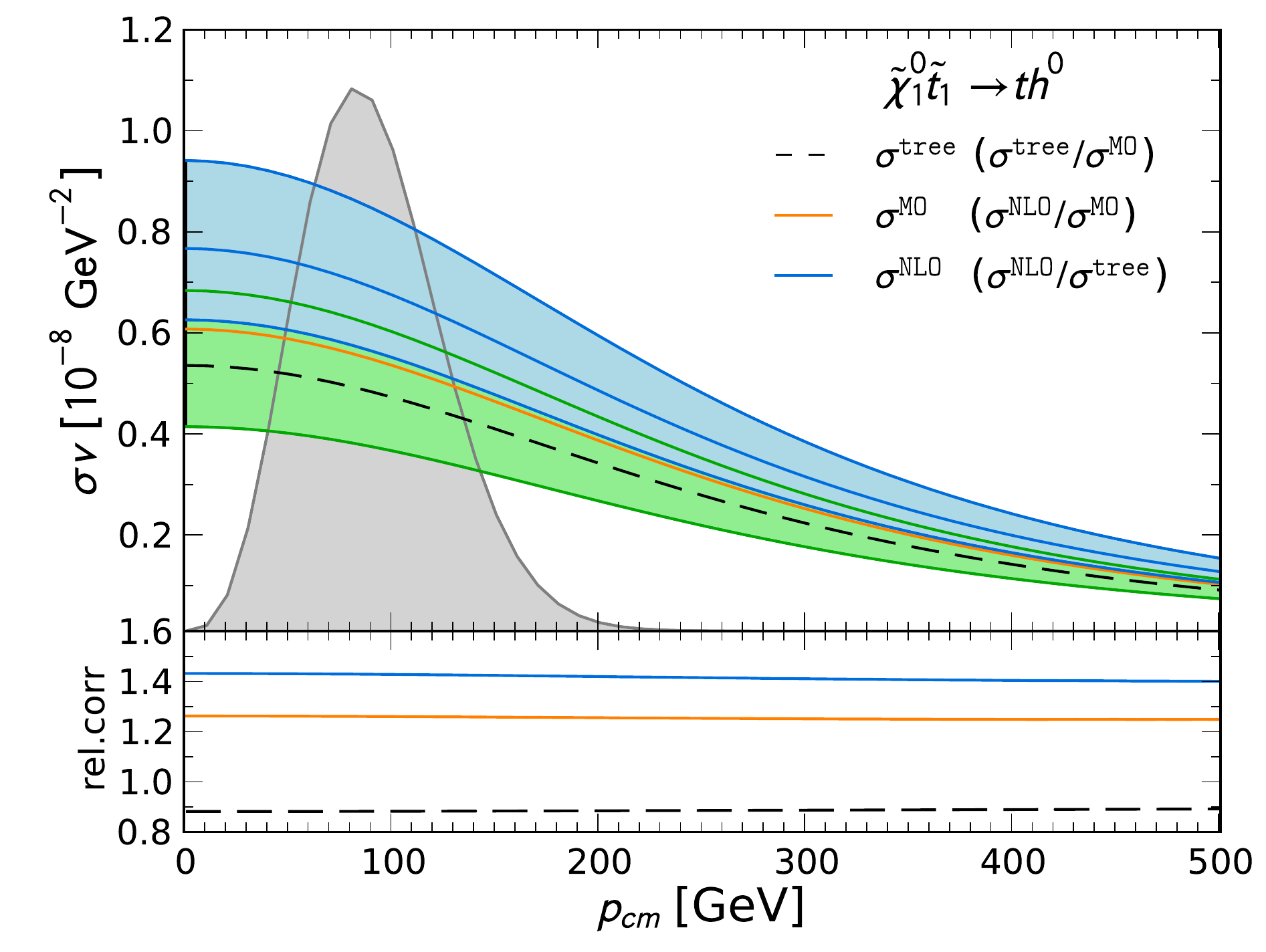}
 \includegraphics[width=0.49\textwidth]{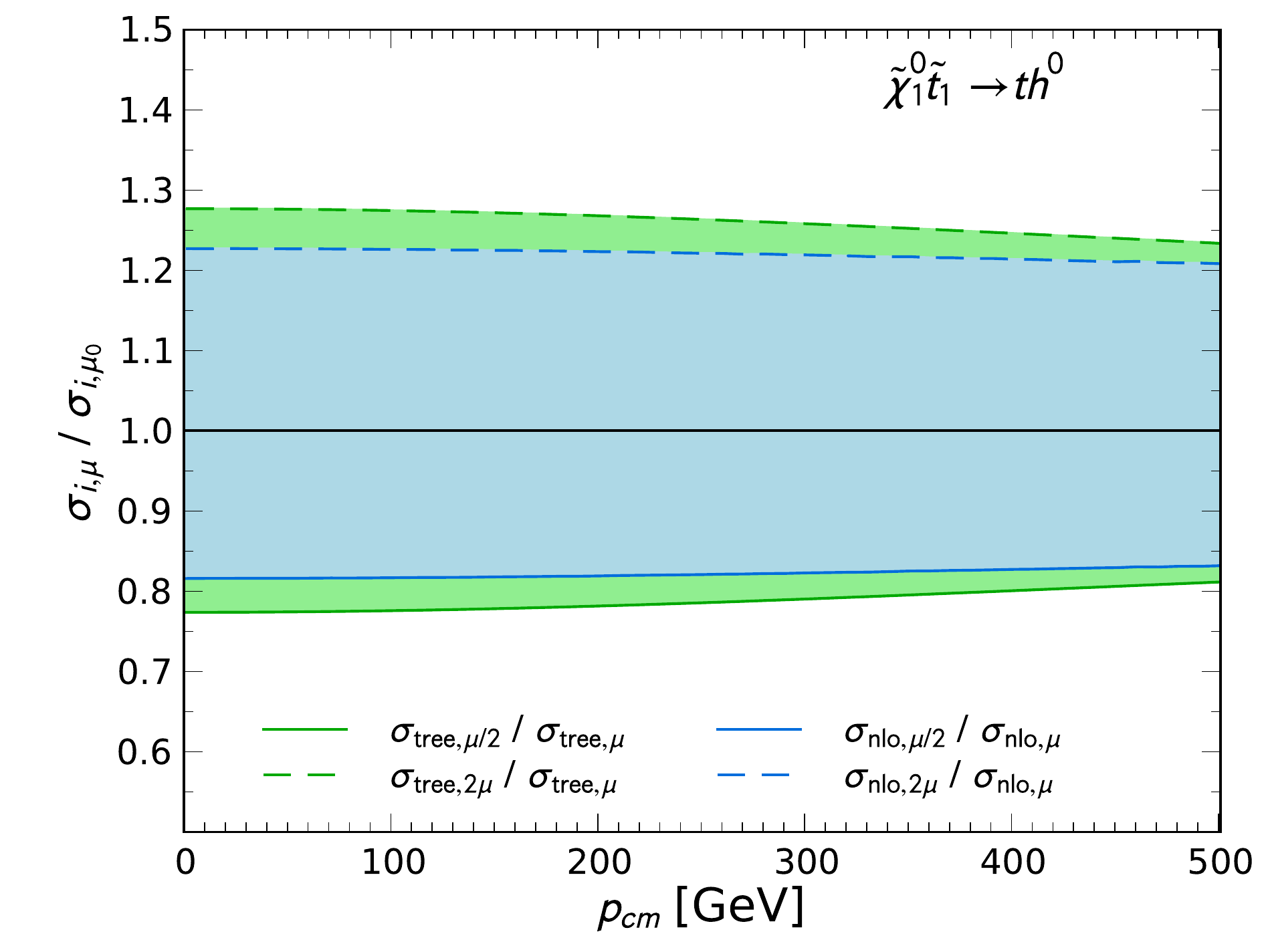}
 \caption{Cross sections of neutralino-stop coannihilation into a top quark and a light neutral
 Higgs boson as a function of the center-of-mass momentum $p_{\rm cm}$ in scenario C. The left
 plots show the annihilation cross sections calculated at the tree-level (black dashed lines)
 and at the one-loop level (blue solid lines) including the corresponding uncertainties from
 variations of the renormalization scale $\mu_R$ by a factor of two around the central scale
 (green and blue shaded bands). We also indicate the values obtained with \MO\ (orange solid
 lines). The right plots show the cross sections normalized to their values obtained with the
 central renormalization scale $\mu_R^{\rm central}=1$ TeV. The two lower plots have been obtained
 using the running MSSM $\overline{\rm DR}$ top-quark mass instead of the pole mass.}
 \label{fig:coannih}
\end{figure*}

\begin{figure}
 \includegraphics[width=0.49\textwidth]{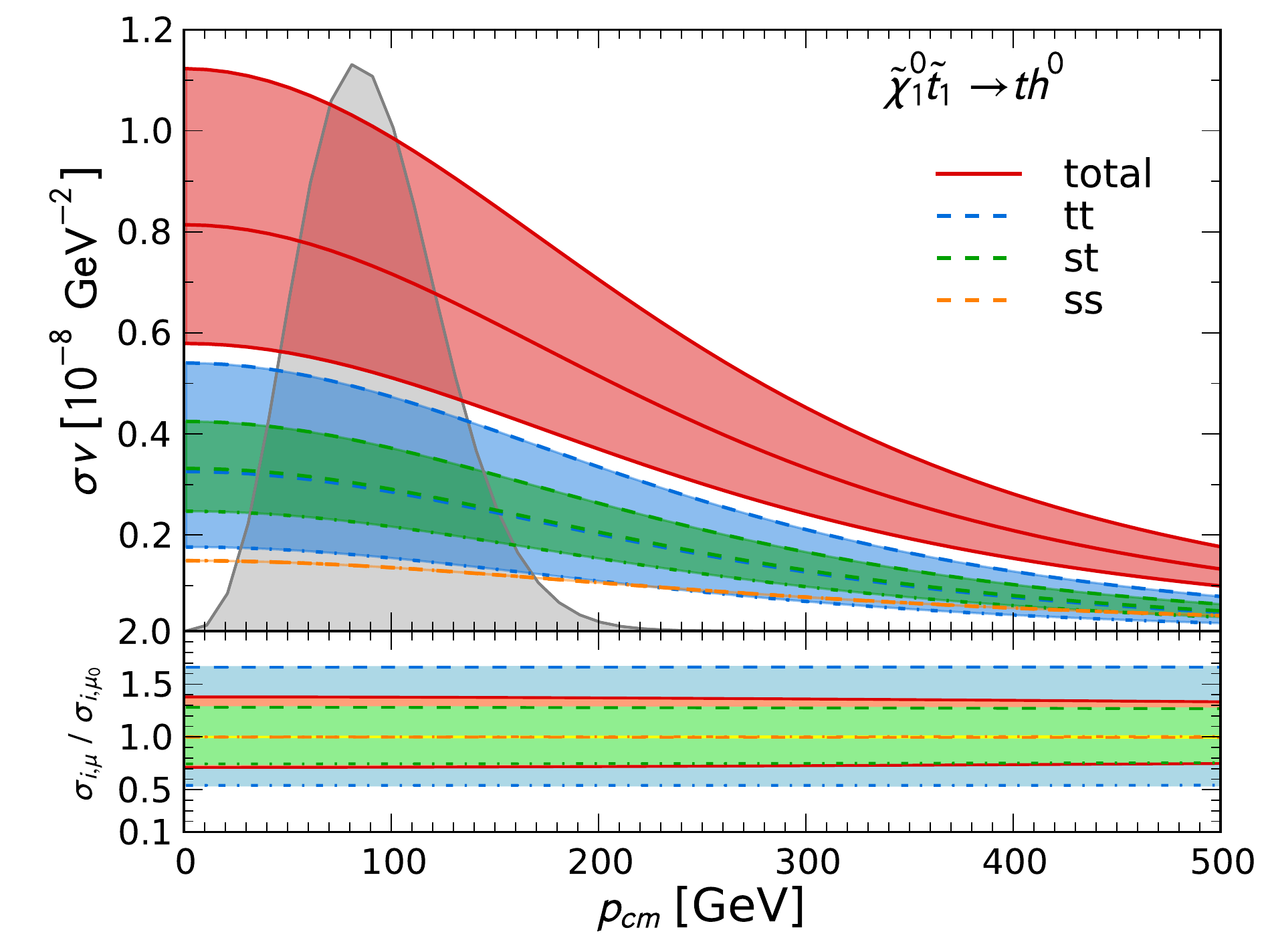}
 \caption{Cross sections of neutralino-stop coannihilation into a top quark and a light neutral
 Higgs boson as a function of the center-of-mass momentum $p_{\rm cm}$ in scenario C. Apart from
 the total cross section, we also show the contributions from the individual channels and their
 tree-level uncertainties induced by variations of $A_t$ and $A_b$.}
 \label{fig:coannibreakdownAtb1}
\end{figure}

\begin{figure}
 \includegraphics[width=0.49\textwidth]{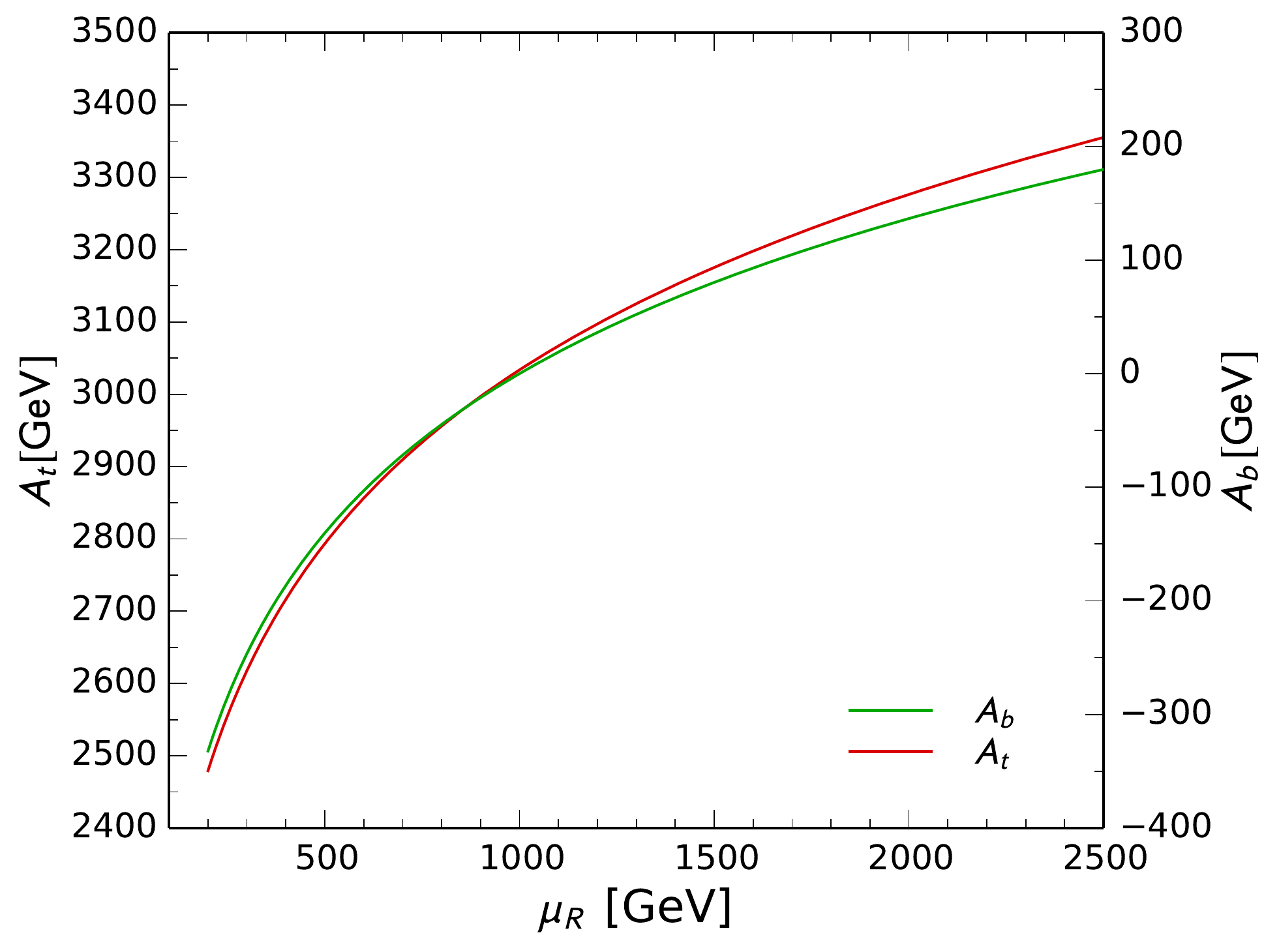}
 \caption{Dependence of the trilinear couplings $A_t$ (red line, left ordinate) and $A_b$ (green
 line, right ordinate) on the renormalization scale $\mu_R$ in scenario C.}
 \label{fig:parametersC1}
\end{figure}

\begin{figure*}
 \includegraphics[width=0.49\textwidth]{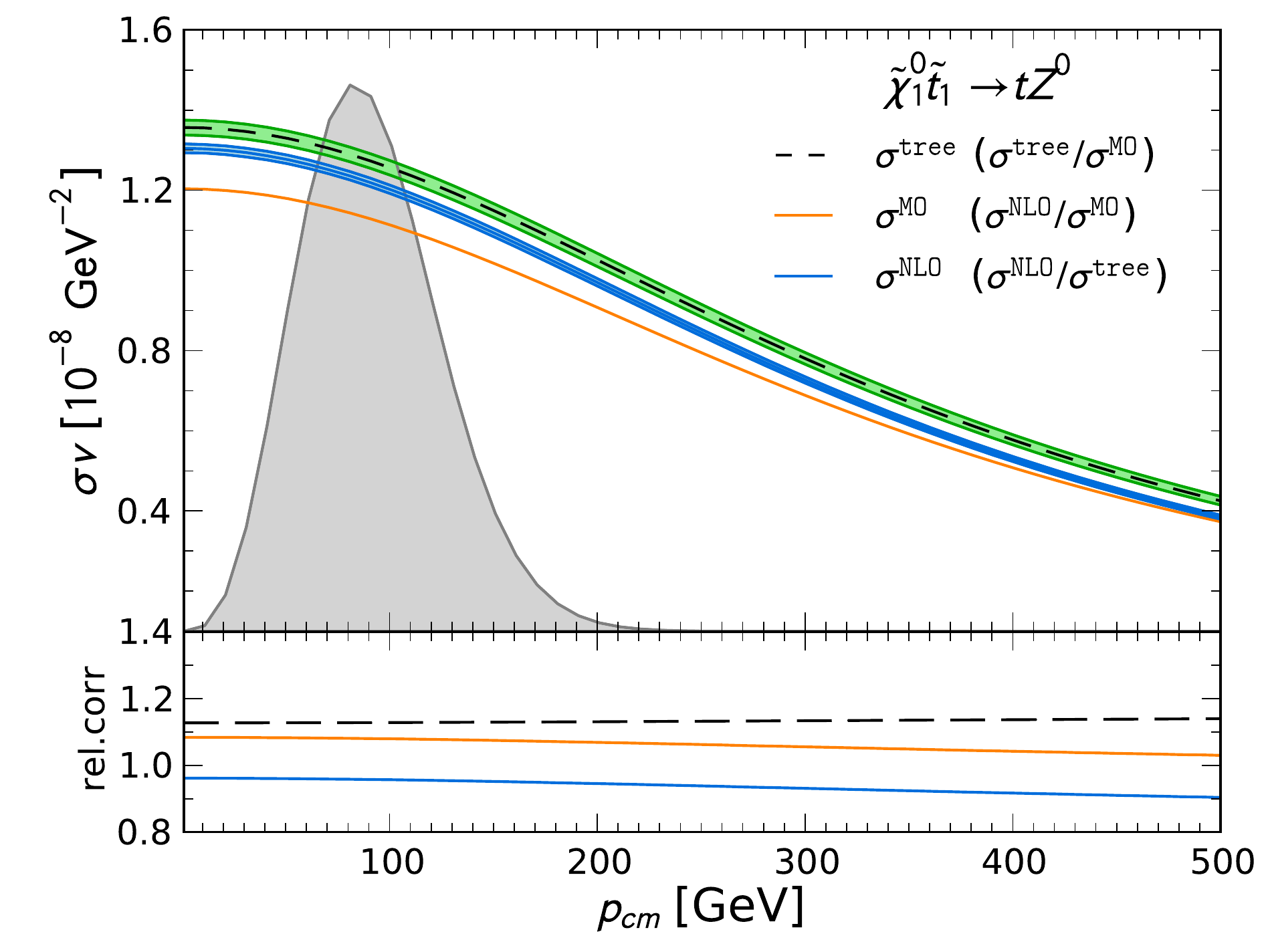}
 \includegraphics[width=0.49\textwidth]{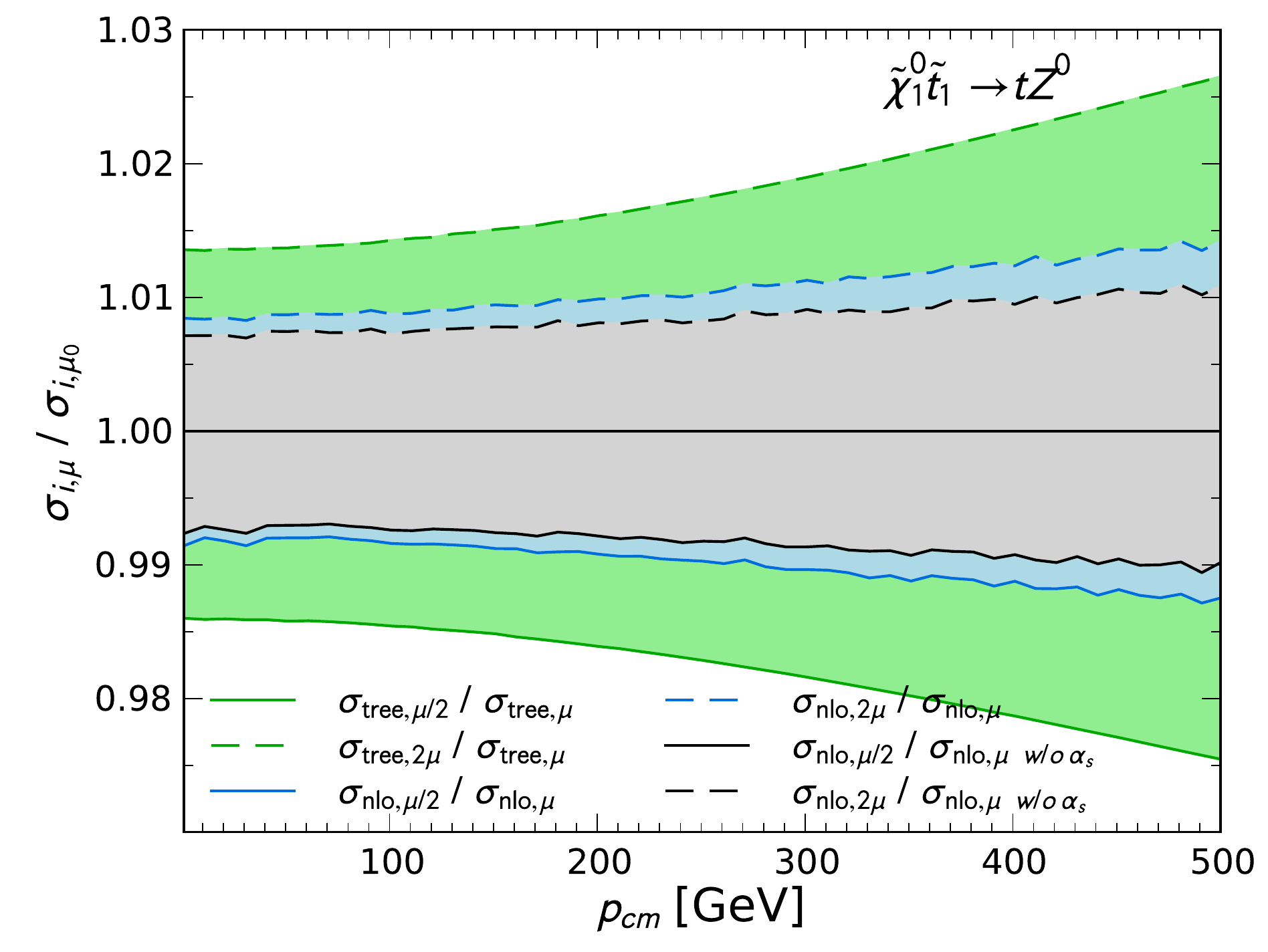}\\
 \caption{Cross sections of neutralino-stop coannihilation into a top quark and a $Z^0$ boson as
 a function of the center-of-mass momentum $p_{\rm cm}$ in scenario C. The left plot shows the
 annihilation cross section calculated at the tree-level (black dashed line) and at the one-loop
 level (blue solid line) including the corresponding uncertainties from variations of the
 renormalization scale $\mu_R$ by a factor of two around the central scale (green and blue shaded
 bands). We also indicate the values obtained with \MO\ (orange solid line). The right plot shows
 the cross sections normalized to their values obtained with the central renormalization scale
 $\mu_R^{\rm central}=1$ TeV, at NLO also without varying the renormalization scale $\mu_R$ in the
 strong coupling constant $\alpha_s$ (grey shaded band).}
 \label{fig:coanniZ}
\end{figure*}

\begin{figure}
 \includegraphics[width=0.49\textwidth]{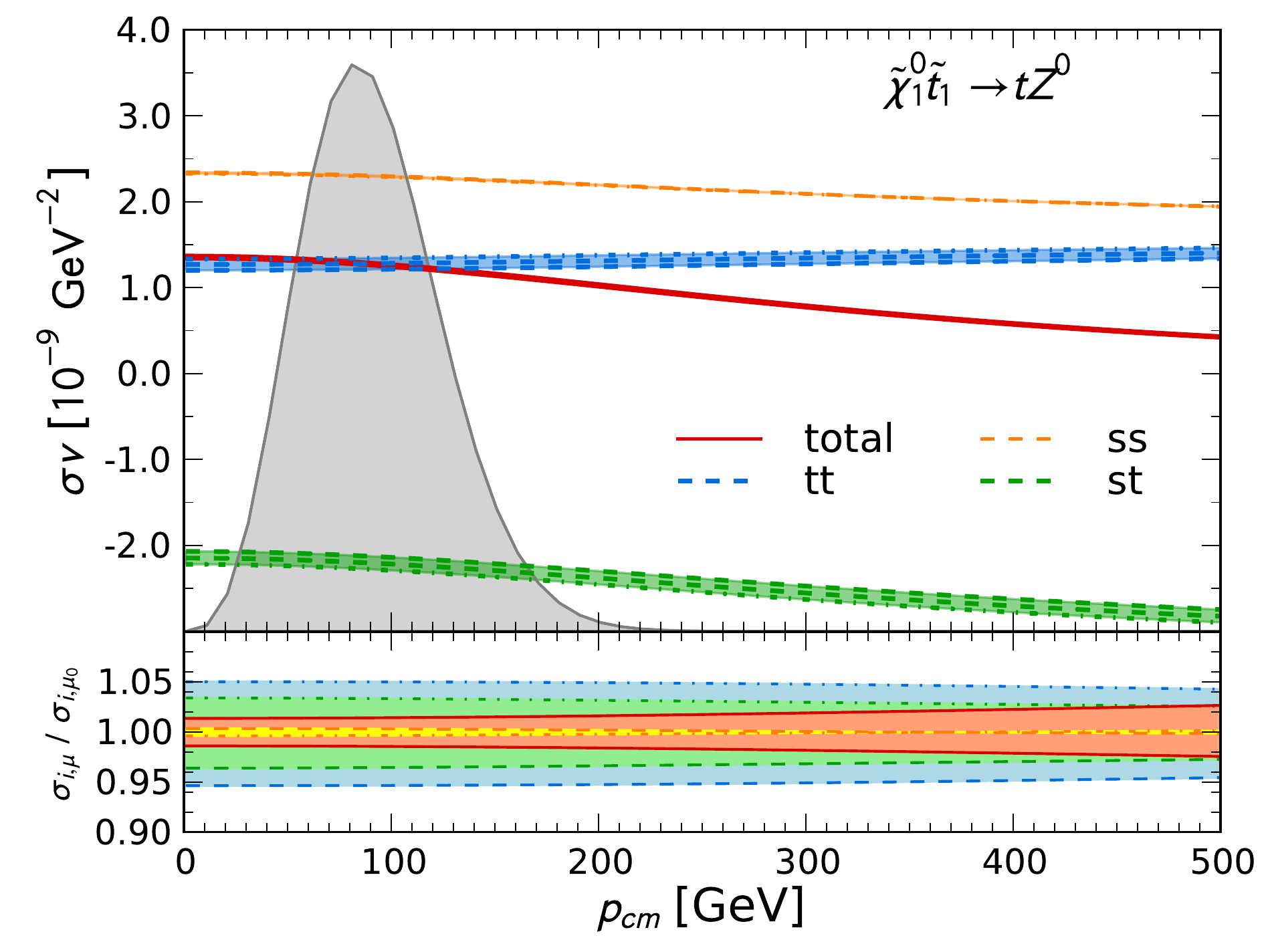}	
 \caption{Cross sections of neutralino-stop coannihilation into a top quark and a $Z^0$ boson as
 a function of the center-of-mass momentum $p_{\rm cm}$ in scenario C. Apart from
 the total cross section, we also show the contributions from the individual channels and their
 tree-level uncertainties induced by variations of $A_t$ and $A_b$.}
 \label{fig:coannibreakdownAtb2}
\end{figure}

\begin{figure}
 \includegraphics[width=0.49\textwidth]{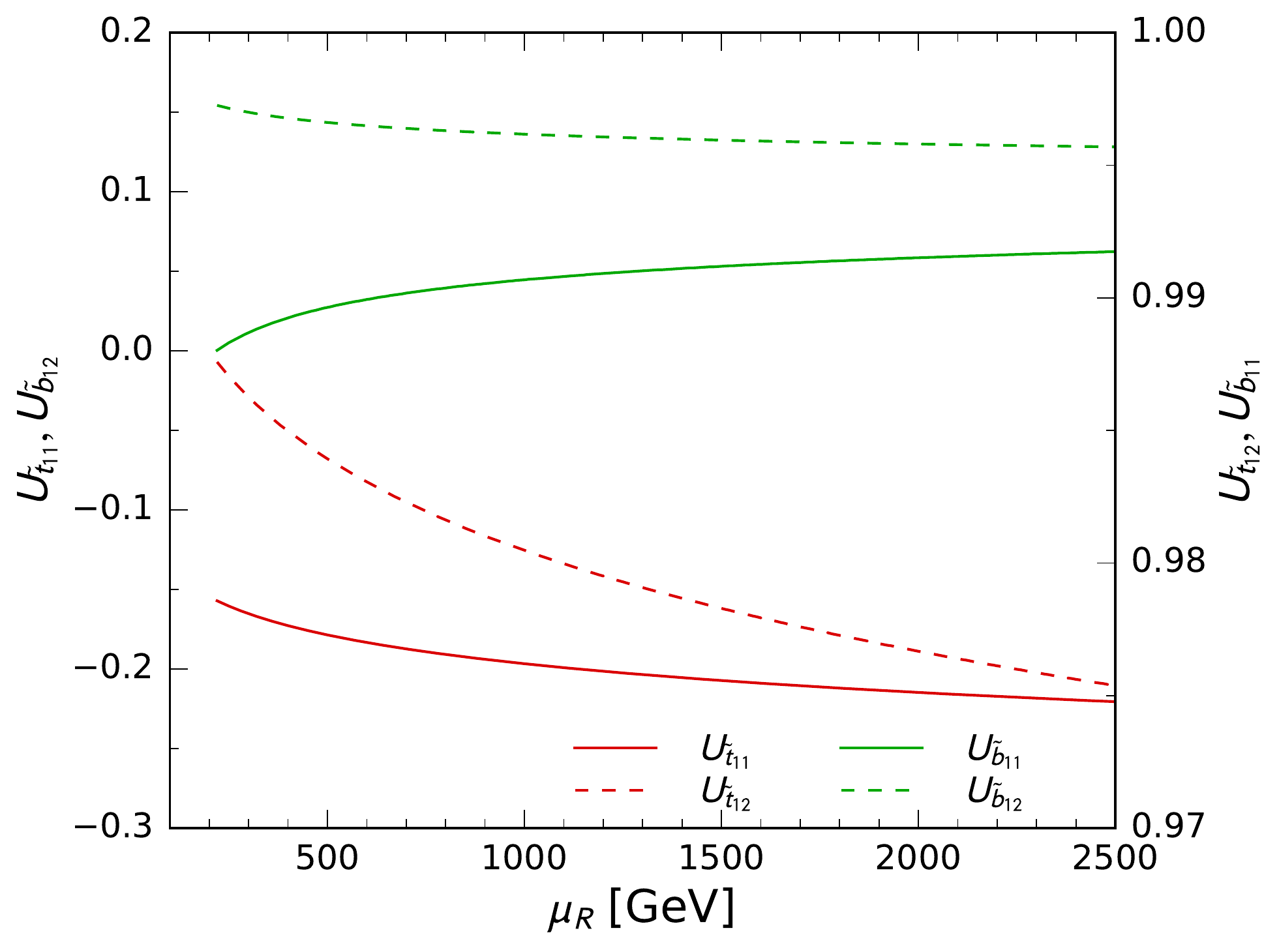}
 \caption{Dependence of the top and bottom squark mixing matrix elements
 on the renormalization scale $\mu_R$ in scenario C.}
 \label{fig:parametersC2}
\end{figure}

\begin{figure*}
 \includegraphics[width=0.49\textwidth]{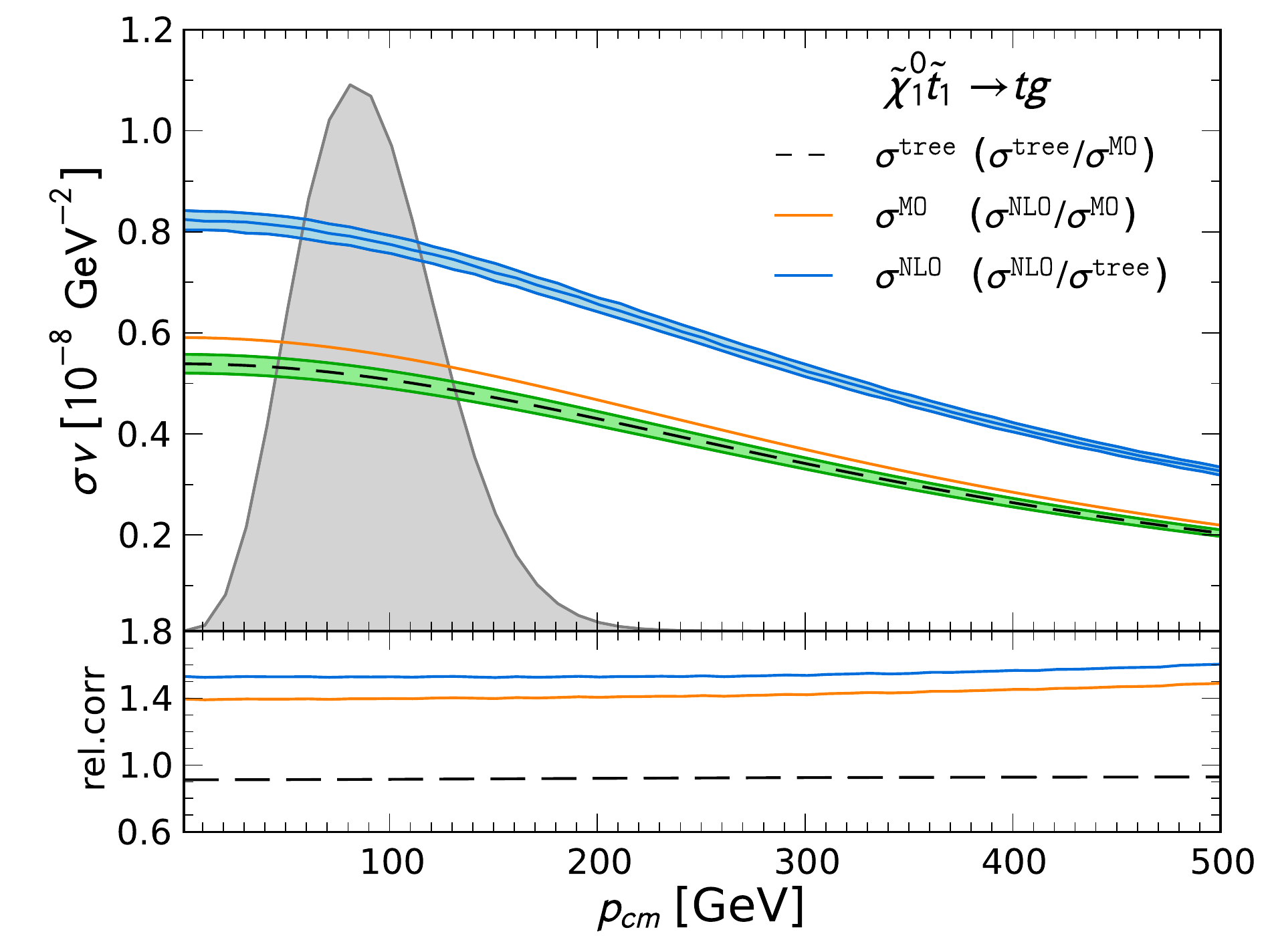}
 \includegraphics[width=0.49\textwidth]{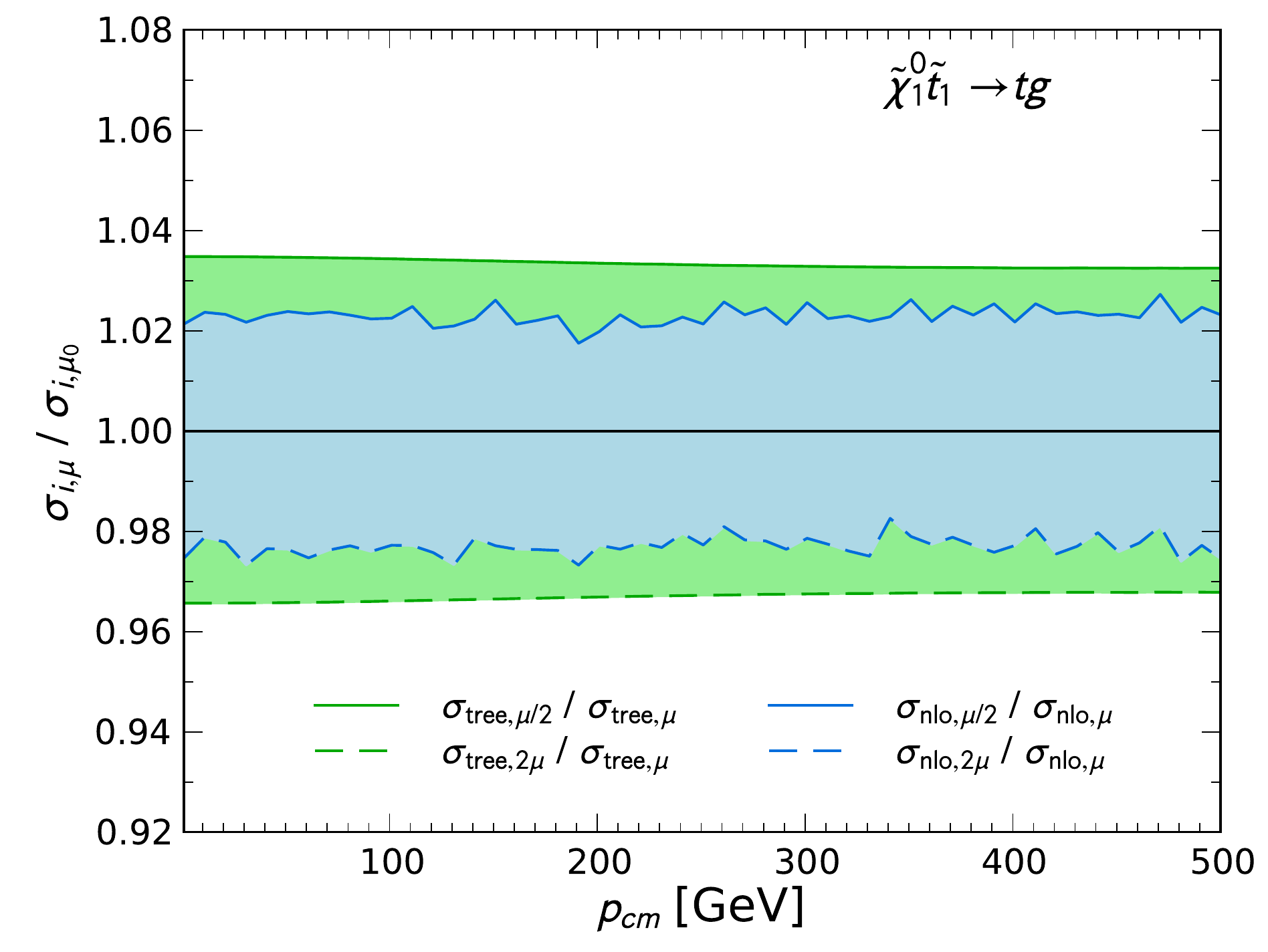}	
 \caption{Same as Fig.\ \ref{fig:coanniZ}, but for neutralino-stop coannihilation into a
 top quark and a gluon in scenario C.}
 \label{fig:coannig}
\end{figure*}

When the lightest neutralino and the lightest stop are almost degenerate in mass, neutralino-stop
coannihilation can contribute dominantly to the total annihilation cross section $\sigma_{\rm ann}$.
Besides leading then to the correct experimentally determined value of the relic density, light
stop scenarios are also well motivated by a Higgs mass of $125.09 \pm 0.21 \mathrm{(stat.)} \pm
0.11 \mathrm{(syst.)}~$GeV \cite{Aad:2015zhl} (cf.\ our discussion in Ref.\ \cite{Harz:2012fz}) as
well as by electroweak baryogenesis \cite{Delepine:1996vn}. Since neutralino-stop coannihilation
processes show a different dependence on the renormalization scale from the processes discussed
above, particularly in the case of a top quark and a gluon in the final state, a dedicated
analysis of this situation is in order.

In the following analysis, we focus on a benchmark scenario that has already been discussed in
Ref.\ \cite{Harz:2014tma}. It involves all processes for which we provide SUSY-QCD corrections and
is thus exemplary. In particular, it features important contributions of neutralino-stop
coannihilation into a top quark and a gluon ($23\%$) or a Higgs boson ($23\%$), but also of
neutralino pair annihilation into top quarks ($16\%$) as well as stop-antistop annihilation
($10\%$) into electroweak final states. The full list of (co)annihilation channels is given in
Tab.\ \ref{ScenarioChannels}. Now that the Sommerfeld enhancement and the full one-loop
corrections for stop-antistop annihilation are available \cite{Harz:2014gaa}, we can extend our study
of this scenario in terms of the total correction to the relic density and its scale dependence.
Distinctive for this scenario are low neutralino and stop masses of $338.3$ GeV and $376.3$ GeV,
respectively, with a small mass difference of only $38$ GeV. This allows not only for
neutralino-stop coannihilation, but also for stop-stop annihilation processes. This specific
mass configuration, as well as the dominance of the top-Higgs final state, is in particular
triggered by the closeness of $M_1$ and $M_{\tilde{u}_3}$ as well as a large trilinear coupling
$A_t$. Further key features of this scenario can be found in Tabs.\ \ref{ScenarioList} and
\ref{ScenarioProps}.

First, we focus on one of the two leading processes, $\tilde{\chi}_{1}^0\tilde{t}_1\to t h^0$, for
which we show the absolute cross sections as a function of the center-of-mass momentum
$p_{\rm cm}$ in the upper left plot of Fig.\ \ref{fig:coannih}. A distinct shift between the
leading-order result calculated with \MO\ (orange solid line) and our tree-level calculation
(black dashed line) is clearly visible. This difference is again caused by different treatments
of the top-quark mass. Whereas we use the pole (on-shell) mass $m_t^{\mathrm{OS}}$ throughout our
calculation, \MO\ uses an an effective top-quark mass
\begin{align}
 m_{\mathrm{eff}}^2&(\mu_R) ~=~ m_t^2(\mu_R) \biggr[ 1 + 5.67 a +
 \big( 35.94 - 1.36 n_f \big) a^2 \nonumber \\ 
 &+ \Big[ 164.14 - n_f \big( 25.77 - 0.259 n_f \big) \Big] a^3  \biggr]
 \label{eq:mteff}
\end{align}
with $a = \alpha_s(\mu_R)/\pi$, defined according to the {\tt SLHAplus} library
\cite{Belanger:2010st}. $m_t(\mu_R)$ and $\alpha_s(\mu_R)$ are the running
top-quark mass and strong coupling constant in the SM $\overline{\mathrm{MS}}$ scheme,
respectively. As long as $\mu_R > 2 m_t^{\rm OS} = 348.2$ GeV, this effective mass is used for
calculations in \MO. Since the scale is fixed at $\mu_R = 2m_{\chi_1^0} =  676.6$ GeV in \MO,
$m_{\rm eff}$ is therefore used there in our scenario C.

For further insight, we change in the lower left plot
of Fig.\ \ref{fig:coannih} our usual on-shell top-quark mass $m_t^{\rm OS}$ to the MSSM
$\overline{\mathrm{DR}}$ mass $m_t^{\overline{\rm DR}}$, that is also easily available in our
code and that is much closer to the SM $\overline{\rm MS}$ mass than the pole mass.
This corresponds to a change of the renormalization scheme, which should in principle be discussed
at the same level as changes of the renormalization scale.
As expected, the two tree-level calculations differ then much less than before. In addition,
the \MO\ tree-level calculation lies between our leading-order calculation and our NLO results
as expected, since the effective top mass in Eq.\ \eqref{eq:mteff} includes higher-order
corrections to the SM $\overline{\mathrm{MS}}$ top mass. 
While in the upper left plot of Fig.\ \ref{fig:coannih} the \MO\ result lies outside our
tree-level uncertainty (green shaded) band, it falls within it and approaches our NLO
uncertainty (blue shaded) band after the adjustment of our top quark mass in the lower left plot.
When comparing the upper- and lower-left tree-level (green shaded) uncertainty bands, we notice
a significant reduction when we use $m_t^{\overline{\rm DR}}$, which is induced by a now complementary
scale dependence in the running top quark mass and stop mixing angle. The impact of the different
definition of the top-quark mass is particularly enhanced due to the large trilinear coupling
$A_t$ in scenario C and a dominant contribution of the $t$-channel diagram, see Fig.\
\ref{fig:coannibreakdownAtb1}, which depends directly on the top mass through the
squark-squark-Higgs coupling. Nevertheless, we stick to
our choice of an on-shell top mass in the following, as it better fits our supersymmetric
processes and on-shell top-quark final states. As one can see from the NLO scale uncertainty
(blue shaded) bands, which do (hardly) overlap in the upper (lower) left plots, as well as from
two left lower subplots, this choice leads to enhanced perturbative stability and a reduction of
the $K$-factor from 1.4 to less than 1.1.

With the pole mass, the scale dependence is reduced from about $\pm30\%$ at LO to about $\pm
20\%$ at NLO, as one can see in the upper right plot of Fig.\ \ref{fig:coannih}. At LO, the scale
dependence is mainly induced by the large trilinear coupling with a total scale dependence of
about $15\%$ as shown in Fig.\ \ref{fig:parametersC1}. This dependence is reduced at NLO due
to cancellations between the vertex corrections to the squark-squark-Higgs coupling and the
corresponding contributions in the renormalization group equation (RGE) of $A_t$.
The scale dependence of $\alpha_s$, that enters first at NLO, is of minor importance.

Fig.\ \ref{fig:coannibreakdownAtb1} shows a detailed breakdown of the tree-level processes
$\tilde{\chi}_{1}^0 \tilde{t}_1 \to t h$, obtained with variations of $\mu_R$ in the trilinear
coupling $A_t$. It is dominated by the highly scale-dependent $t$-channel, which induces most of
the LO scale dependence in Fig.\ \ref{fig:coannih}. In contrast, the $s$-channel shows almost
no scale dependence. As this channel gains in importance for a larger center-of-mass momentum,
the overall scale dependence diminishes for larger $p_{cm}$, as visible in Fig.\ \ref{fig:coannih}.
Note that the true scale uncertainty of our calculation is likely to be smaller than just
explained. This is due to the fact that we take the value of $A_t (\mu_R)$ directly from
{\tt SPheno}, which includes scale-dependent electroweak loop contributions. These are of
course not cancelled by our SUSY-QCD NLO calculation.

Second, we study in Fig.\ \ref{fig:coanniZ} the process $\tilde{\chi}_{1}^0 \tilde{t}_1 \to t Z^0$.
With 5\%, it contributes subdominantly to the relic density in scenario C, but it is nevertheless
theoretically interesting, as it shows a much smaller scale dependence than the process with a
Higgs final state. The left plot in Fig.\ \ref{fig:coanniZ} shows again a distinct difference
between the two tree-level calculations. In the case of a $Z^0$ boson in the final state, these
differences arise now mostly from differently defined squark mixing angles and the dependent mass
of the heavier stop \cite{Harz:2012fz}. Our NLO calculation results in a correction of about $5\%$
with respect to our tree-level calculation and of about $10\%$ in comparison to \MO, as the lower
left subplot shows.

The right plot of Fig.\ \ref{fig:coanniZ} exhibits more clearly the relative scale dependence of
our LO and NLO calculations. Whereas the former increases with $p_{\rm cm}$ from $\pm1.5\%$ to $\pm
2.5\%$, it always stays below $\pm1.5\%$ at NLO. The grey shaded band has been obtained without
varying $\mu_R$ in $\alpha_s$. As one can see, the implicit scale dependence induced at NLO by
the strong coupling constant is again negligible compared to the loop corrections reducing the
scale dependence of the trilinear coupling.

The increase of the scale dependence with $p_{cm}$ can be understood from Fig.\
\ref{fig:coannibreakdownAtb2}. It shows that the $t$-channel process contributes dominantly to
the scale dependence, as both couplings and the propagator contain scale-dependent parameters,
i.e.\ the squark mixing matrices and the stop mass $m_{\tilde{t}_2}$. As for higher $p_{cm}$ the
stop mass in the propagator is less important, the scale dependence decreases. In contrast, the
$s$-channel contains only the relatively light top-quark propagator and a single source of scale
dependence in the neutralino-stop-top coupling. Overall, it is as important as the $t$-channel,
but less scale dependent. Its scale dependence increases with $p_{\rm cm}$ due to the interplay
between the mixing angles (see Fig.\ \ref{fig:parametersC2}), and this translates into
a slightly increased scale dependence of the total cross section at larger $p_{cm}$.

The scale dependence of the two neutralino-stop coannihilation processes discussed so far was
large with about $\pm 20\%$ at NLO for Higgs final states and very small for the $Z^0$-boson
final state with around $\pm 1.5\%$. In both cases a reduction was visible from LO to NLO
despite the fact that these were purely electroweak processes at tree level.
As a third coannihilation process, we now consider the semi-weak process $\tilde{\chi}_{1}^0
\tilde{t}_1 \to t g$, which contributes again with 23\% to the relic density in scenario C.
As one can see in the left plot of Fig.\ \ref{fig:coannig}, the \MO\ and our tree-level
calculations differ in this case by less than 10\%. However, our NLO corrections induce a
relatively large $K$-factor of more than 1.5.
The right plot in Fig.\ \ref{fig:coannig} shows that the LO scale uncertainty, induced mainly
by the presence of $\alpha_s(\mu_R)$ already at this order, is smaller than $\pm4\%$. This is
due to the fact that $\mu_R=0.5...2$ TeV is already quite large and the running of $\alpha_s$
relatively slow (cf.\ Fig.\ \ref{fig:ScaleDependence}). The scale dependence is reduced by the
NLO corrections as expected to a level of $\pm2\%$.

\begin{figure*}
 \includegraphics[width=0.49\textwidth]{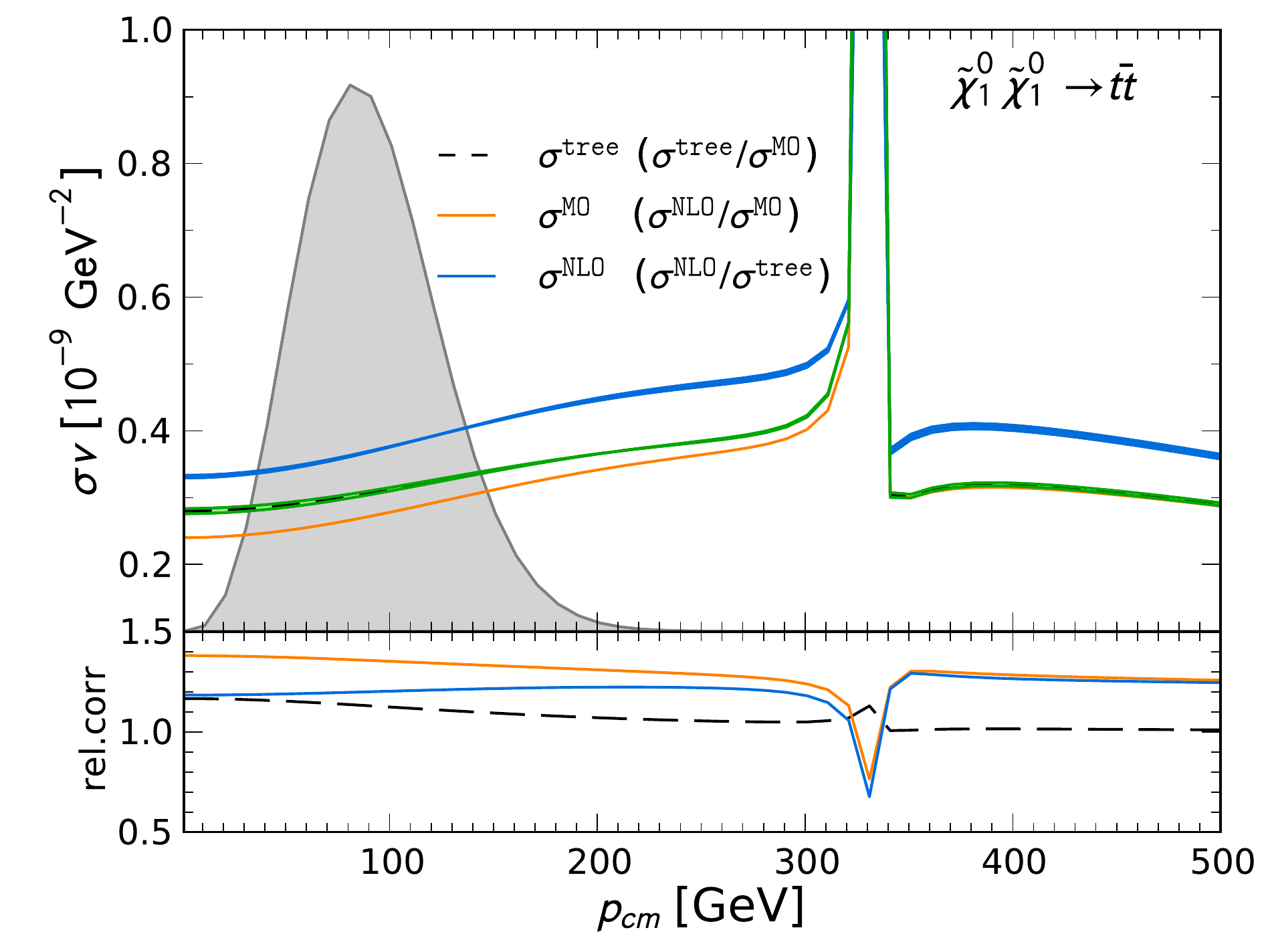}
 \includegraphics[width=0.49\textwidth]{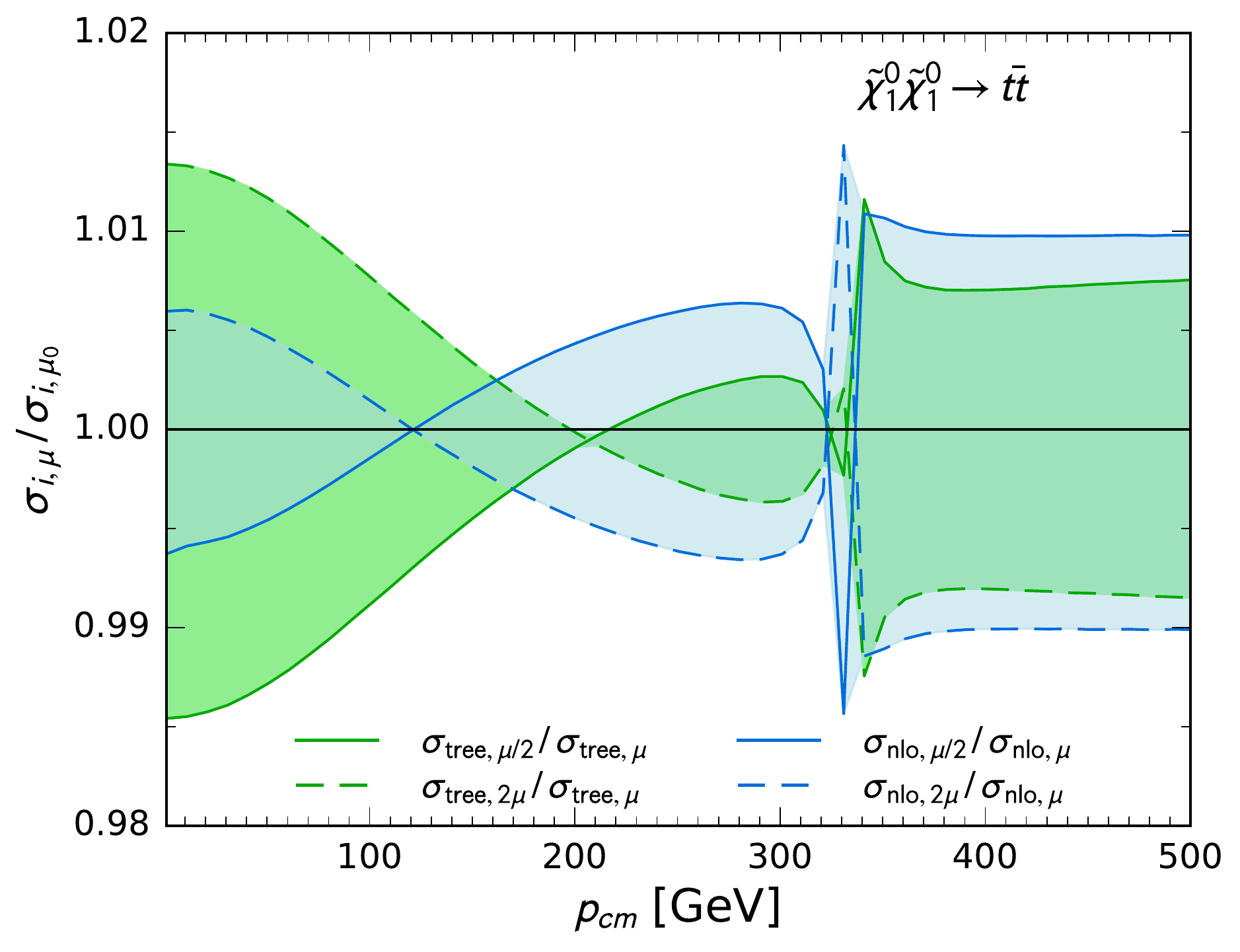}
 \caption{Same as Fig.\ \ref{fig:coanniZ}, but for neutralino annihilation into a pair of top
 quarks in scenario C.}
 \label{fig:anni}
\end{figure*}

The third most important process in scenario C contributing 16\% to the relic density is the
pair annihilation of two neutralinos into a top-antitop pair, $\tilde{\chi}^0_1 \tilde{\chi}^0_1
\to t \bar{t}$. This is again a purely electroweak process at LO and is similar to the processes
discussed for scenario A in Sec.\ \ref{Sec:GauginoCoAnni}. In the left plot of Fig.\
\ref{fig:anni} we show the annihilation cross section of this channel as a function of the
center-of-mass momentum $p_{\rm cm}$. It exhibits a peak at $p_{\rm cm}\simeq 330$ GeV from the
heavy neutral Higgs resonances, that lies however far beyond the peak of the thermal dark
matter velocity distribution. At and below this peak, the \MO\ and our tree-level calculations
differ again visibly due to the different mass and mixing angle definitions, but agree rather
well at high energy. The NLO $K$-factor is about 1.25 except at the resonance.

The dependence on the renormalization scale is, however, less important in this case, as can be
seen from the right plot of Fig.\ \ref{fig:anni}. Varying the renormalization scale between
$\mu_R = 0.5$ and $\mu_R = 2$ TeV leads to a variation of the LO cross section of less than two
percent. Since the top-quark mass is treated in the on-shell scheme (see Sec.\ \ref{Technical}),
it is independent of the scale. Furthermore, the scale-dependent trilinear coupling does not
enter the calculation at the tree-level, since the lightest neutralino is a pure bino. We have
verified numerically that the scale-dependence of the heavier stop mass does not influence the
cross section in a visible way. The only relevant scale-dependent parameter is thus the stop
mixing angle $\theta_{\tilde{t}}$, which is relevant for the squark exchange in $t$- and $u$-channel
diagrams. The latter dominate the cross section, except for the $s$-channel resonance of heavy
Higgs bosons around $p_{\rm cm} \approx 330$ GeV. The observed dependence on the scale can thus be
attributed to the variation of the stop mixing angle shown in Fig.\ \ref{fig:parametersC2}.

\begin{figure*}
 \includegraphics[width=0.49\textwidth]{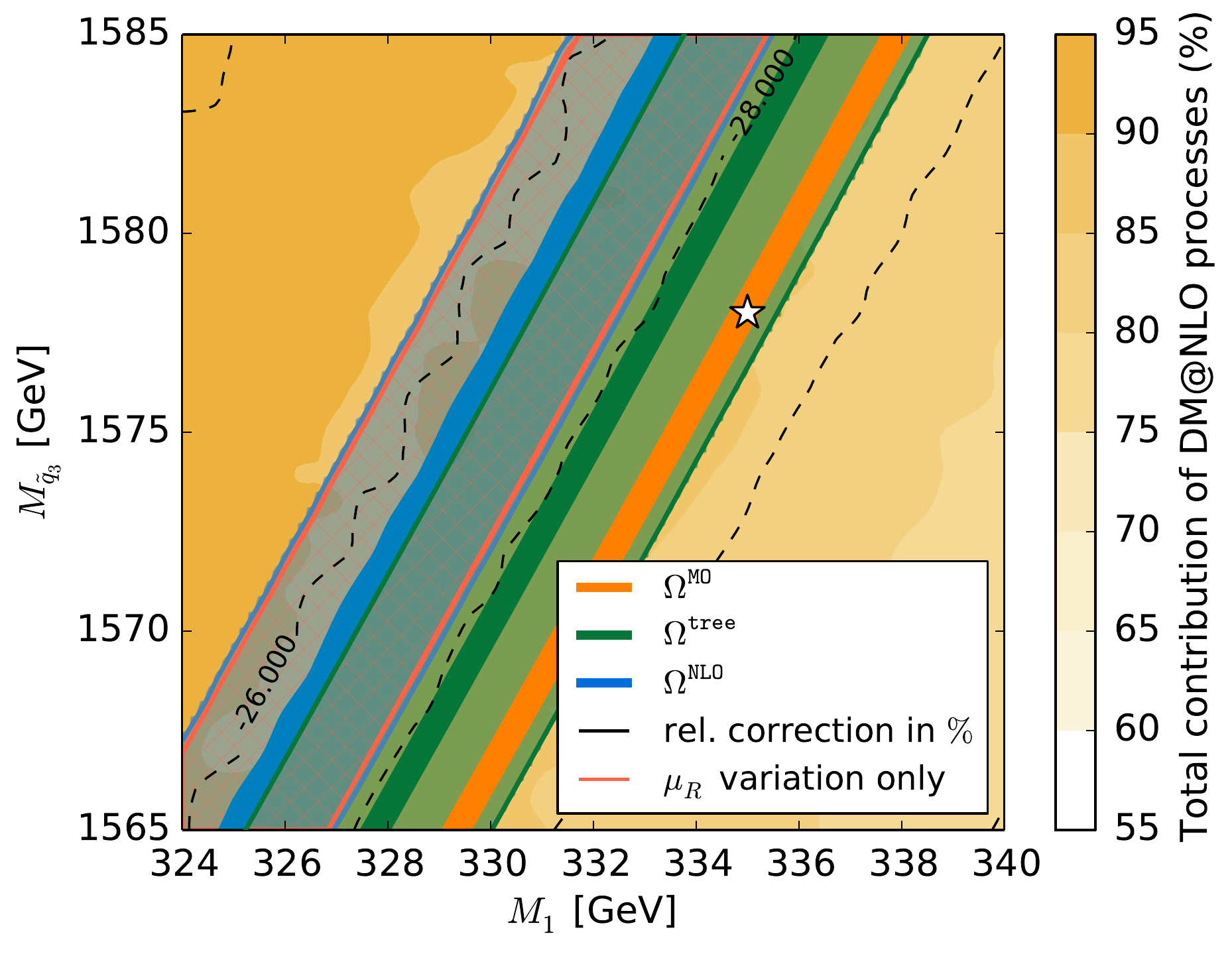}
 \includegraphics[width=0.49\textwidth]{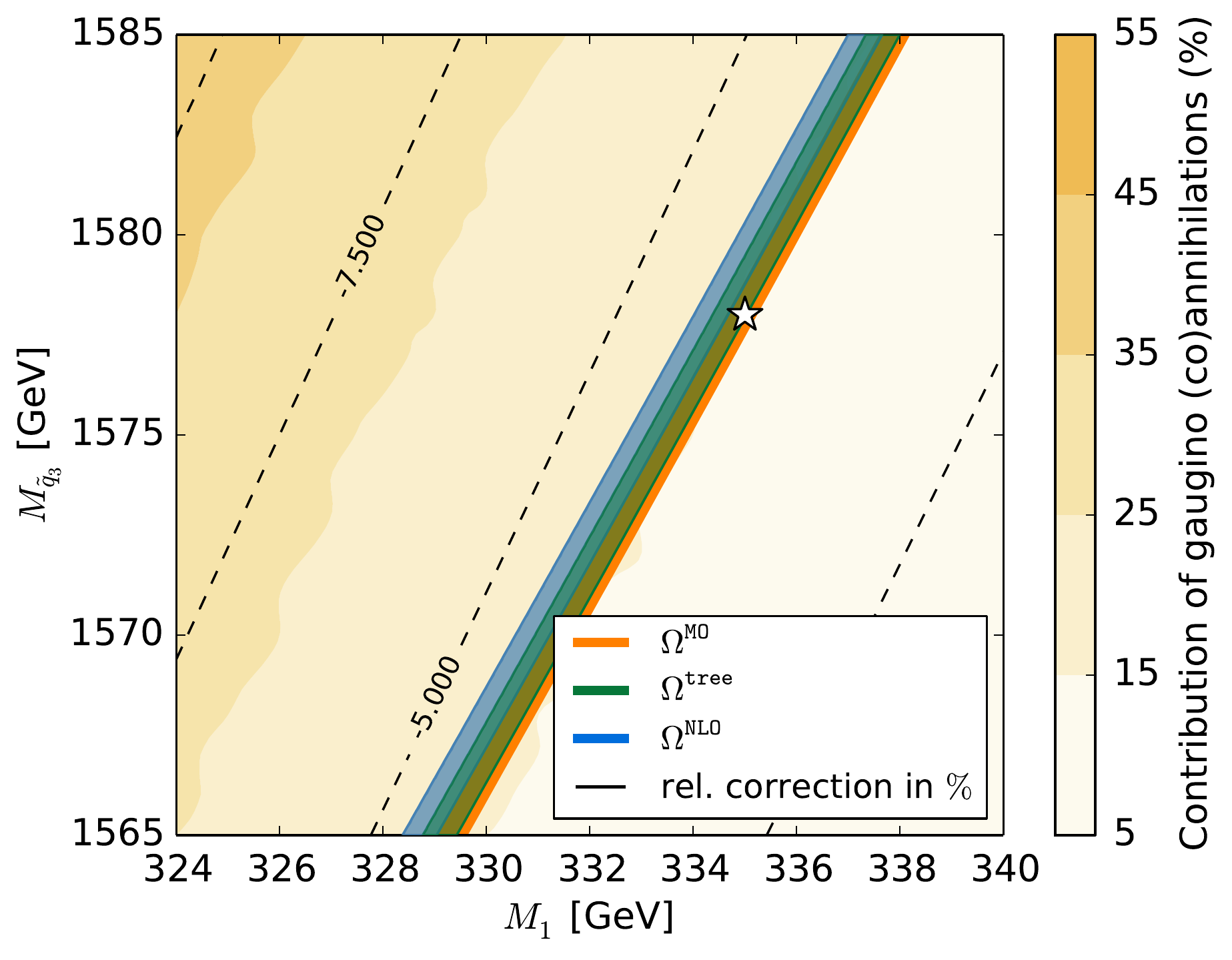}
 \includegraphics[width=0.49\textwidth]{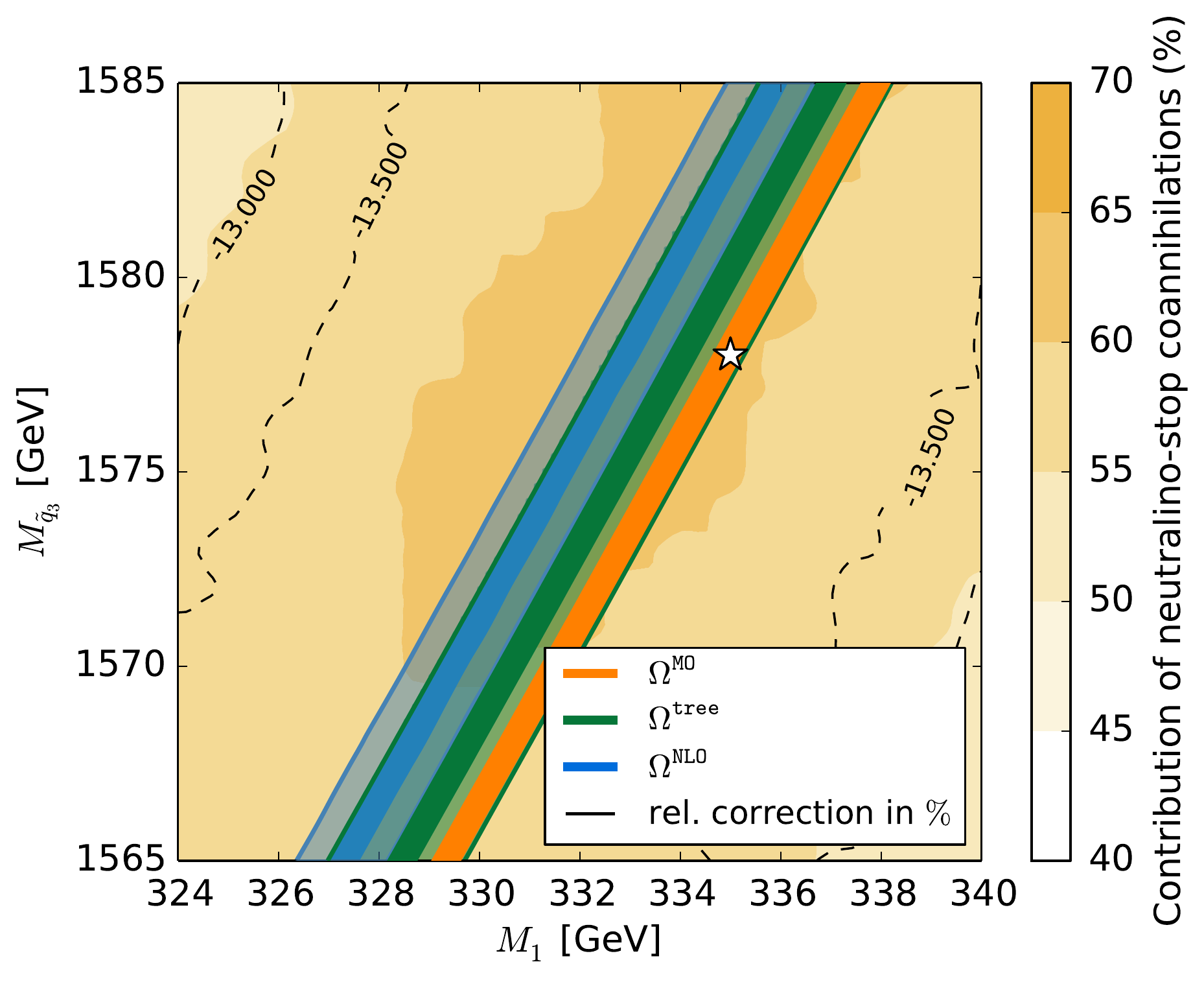}
 \includegraphics[width=0.49\textwidth]{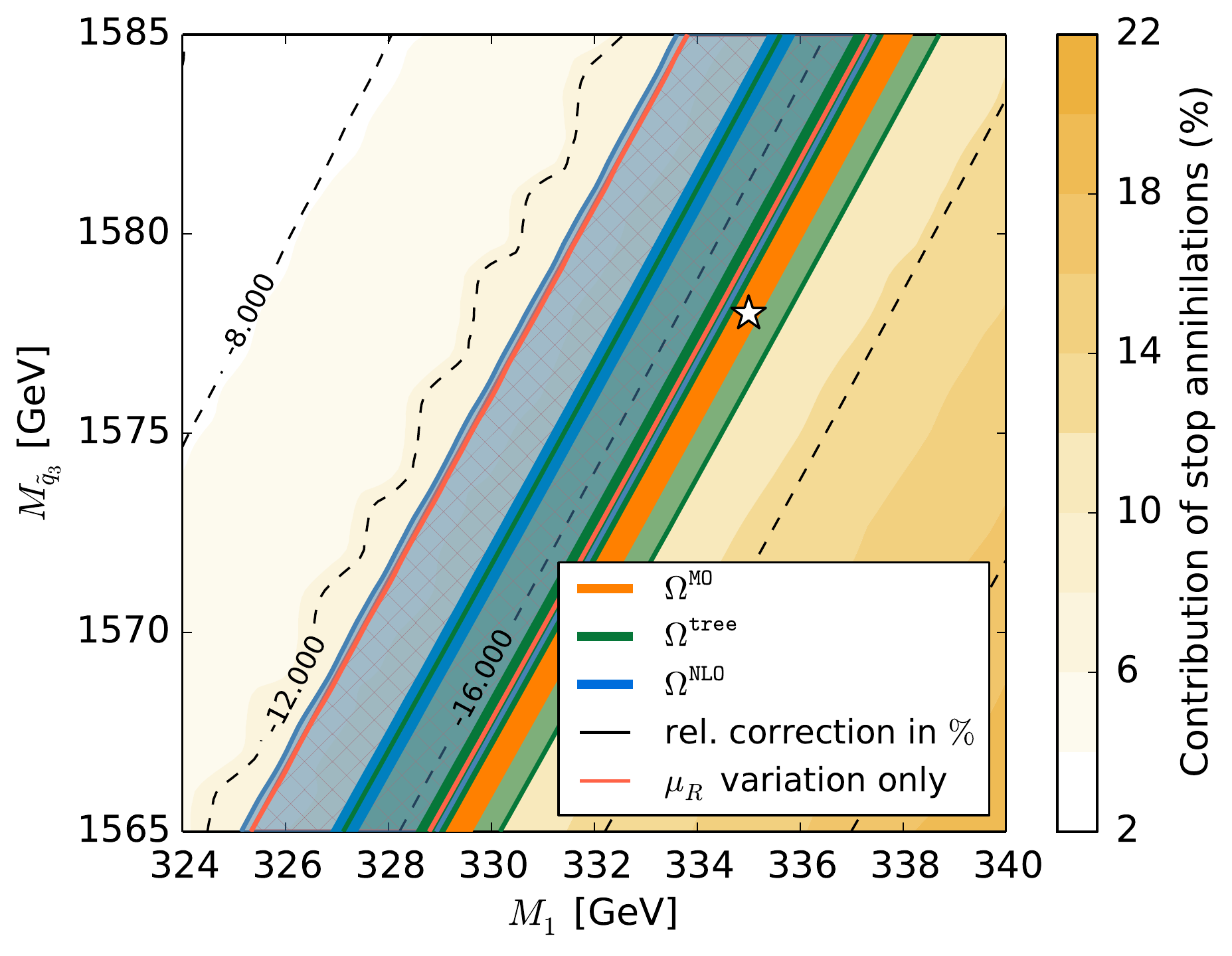}
 \caption{Cosmologically preferred regions with respect to Eq.\ \eqref{Planck} in the
 $M_1$-$M_{\tilde{q}_3}$-plane surrounding scenario C (white star). The solid bands correspond to
 the limits of Eq.\ \eqref{Planck}, while the shaded bands include an additional variation of the
 renormalization scale $\mu_R$ between $\mu_R^{\rm central}/2$ and $2\mu_R^{\rm central}$. The shades of
 yellow indicate the percentages of the total and individual annihilation cross sections which
 have been corrected in our calculation, while the dashed lines represent contours of the relative
 correction to the relic density in percent.}
 \label{fig:RelicScenC}
\end{figure*}

Let us now turn to the discussion of the neutralino relic density, which we show in Fig.\
\ref{fig:RelicScenC} in the $M_1$-$M_{\tilde{q}_3}$-plane surrounding scenario C using the same
color code as before. The three bands correspond to that part of the parameter space, which
leads to a relic density compatible with the Planck limits of Eq.\ \eqref{Planck} when using
the \MO\ effective tree level (orange), our tree level (green) and our full
$\mathcal{O}(\alpha_s)$ calculation (blue). The two last ones are surrounded by green and
blue shaded bands, which illustrate the scale uncertainty obtained by varying both $\mu_R$ and
$\mu_C$ by factors of two. The red shaded bands correspond to varying only $\mu_R$. The yellow
background contours denote the relative contribution of processes we have implemented at NLO. 

We start our discussion with the upper left plot, where we have used all our subprojects in
parallel. The total contribution of \DMNLO\ processes amounts to 85-90\% in the region of the
three bands. This means that our code is able to improve on almost all the relevant processes in
the selected part of the parameter space. As expected from our previous work, the three central
predictions clearly separate. The impact of the radiative corrections on the resulting relic
density is much larger than the experimental uncertainty given by Planck. More precisely, we
observe a remarkable relative shift of our full result in comparison to \MO\ 
for the relic density
of 27\%. This is indicated by the black dashed lines in Fig.\ \ref{fig:RelicScenC}.
Furthermore we observe rather large theoretical uncertainty bands. Within the scale uncertainty,
our tree level result agrees with \MO\ and our full result, but our full result does not agree
with \MO. Moreover, the scale dependence of the full result is smaller than the scale dependence
of our tree level as expected after our detailed discussion above. The impact of varying $\mu_C$
is subdominant in this case, i.e.\ the blue shaded and red hashed bands almost agree. We return
to this point when discussing the lower right plot of Fig.\ \ref{fig:RelicScenC}.

It is quite illustrative to compare this plot with the right plot of Fig.\ 11 in Ref.\
\cite{Harz:2014tma}, which shows the same part of the parameter space. In comparison, we have
increased the contribution of our subprocesses from 78 to 88\%, enhanced the corrections
to the relic density from 18 to 27\% and thus increased the shift of the compatible relic 
density band. This is mainly due to the fact that stop annihilation into electroweak final states
is now available. In the following, we decompose the upper left plot of Fig.\ \ref{fig:RelicScenC}
into those pertinent to the individual subprocess contributions in order to analyze the origin of
their scale dependence.

In the upper right plot, we have corrected only the gaugino (co)annihilation processes, i.e.\ in
particular $\tilde{\chi}^0_1\tilde{\chi}^0_1\rightarrow t\bar{t}$, which contributes 16\% to the
relic density (see Tab.\ \ref{ScenarioChannels}). Its relative contribution is also illustrated
by the shades of yellow and increases to more than 35\% in the direction of the upper
left corner in the $M_1$-$M_{\tilde{q}_3}$-plane. This is precisely the direction of a bigger mass
gap between the lightest neutralino and the lightest stop, where stop annihilation becomes
irrelevant and neutralino-stop coannihilation less relevant.
The three uncertainty bands overlap and show almost no dependence on the scale. Still the
relative shift of the relic density amounts to 4 to 5\%, even though in scenario C gaugino
(co)annihilation is of minor importance compared to the other subprocesses. If we increase its
contribution, i.e.\ look at the upper left corner of the plot, this process causes a relative
shift of more than 10\% when contributing 35\%, which is quite remarkable for neutralino
annihilation (compare with e.g.\ Ref.\ \cite{Herrmann:2014kma}). This is of course due to the large
correction of the corresponding cross section (see Fig.\ \ref{fig:anni} and the associated
discussion). We remind the reader that SUSY-QCD corrections for neutralino annihilation into
heavy quarks are particularly relevant, when there is no small mass splitting between neutralinos
and stops \cite{Herrmann:2007ku, Herrmann:2009wk, Herrmann:2009mp}.

The lower left plot shows the impact of neutralino-stop coannihilation. Together, these processes
contribute more than 60\% and cause a relative shift of the relic density of 14\%. The scale
dependence is similar to the upper left plot. Within its scale uncertainty our tree level agrees
with \MO\ and our full result, but our full result differs from \MO. This is in agreement with
our discussion of the contributing cross sections (cf.\ Figs. \ref{fig:coannih}, \ref{fig:coanniZ}
and \ref{fig:coannig}). Remember that in particular the process $\tilde{\chi}^0_1\tilde{t}_1
\rightarrow th^0$ featured a large scale dependence caused by its dependence on the trilinear
coupling $A_t$. As this process contributes 23\% to the relic density, the relic density also
shows a rather large scale dependence. Furthermore, the scale dependence of the relic density
decreases at NLO, as it is the case for all the discussed cross sections.

The remaining, lower right plot shows the impact of stop annihilation. It contributes roughly
10\% in the region of the three bands and becomes more important in the direction of the lower
right corner of the $M_1$-$M_{\tilde{q}_3}$-plane, the direction of a smaller mass gap between the
lightest neutralino and the lightest stop. Hence the shades of yellow are complementary to the
contours for gaugino (co)annihilation shown directly above. 
The relative shift of the relic density amounts to roughly 15\%, which is very large for a
subprocess contributing only 10\%. It is caused by the large radiative corrections on the cross
sections induced by Sommerfeld enhancement. Nevertheless, these corrections seem to be
surprisingly large compared to our previous work \cite{Harz:2014gaa} and the previous subsection.
This has the following reason. Remember that the Sommerfeld enhancement becomes relevant for
small relative velocities of the incoming particles. On the other hand, when calculating the
relic density, the cross sections are weighted with the thermal distribution of these particles.
In scenario C we have rather light neutralinos and stops (338.3 GeV and 376.3 GeV, cf. Tab.\
\ref{ScenarioProps}), which let the thermal distribution peak below $p_{\mathrm{cm}}\sim 100$ GeV
(cf.\ Fig.\ \ref{fig:coanniZ}). In scenario B the particles were much heavier (1306.3 GeV and
1361.7 GeV, cf.\ Tab.\ \ref{ScenarioProps}), and the thermal distribution peaked above
$p_{\mathrm{cm}}\sim 250$ GeV (cf.\ Fig.\ \ref{fig:ScaleDependenciesScenB}). This means that the
low-velocity tail of the cross section, where the Sommerfeld enhancement really matters, gets
now a bigger weight than in scenario B \cite{Harz:2014gaa}, and hence the Sommerfeld enhancement affects
the relic density even more drastically.
We also observe rather broad uncertainty bands for our tree-level and full results. The first one
is mainly triggered by the dependence of the tree level on the trilinear coupling $A_t$. The
second one is caused by the remaining dependence on $A_t$ and the dependence of the SUSY-QCD
corrections on $\alpha_s$. This is in agreement with the findings of the previos subsection.
However, in comparison to the previous subsection we also find a difference. The impact of
varying $\mu_C$ is smaller than before, i.e.\ the red and blue shaded bands almost agree.
This is due to the fact that $A_t$ constitutes the dominant source of scale dependence in this
scenario. Note that varying $A_t$ between 0.5 and 2 TeV leads to a change of $\sim15$\% in this
scenario (cf.\ Fig.\ \ref{fig:parametersC1}) instead of $\sim3$\% in scenario B (cf. Fig.\
\ref{fig:ScaleDependenceAt}). This leads to a much broader tree-level uncertainty than before,
and the green shaded band in the lower right plot of Fig.\ \ref{fig:RelicScenC} is larger
than in Fig.\ \ref{fig:relicScenB}. When taking into account all subprocesses in parallel
as in the upper left plot of Fig.\ \ref{fig:RelicScenC}, the impact of $\mu_C$ is further
reduced, since the gaugino (co)annihilation and neutralino-stop coanniliation processes depend
only on $\mu_R$.

An extraction of the pMSSM-11 parameters from the Planck relic
density with \MO\ would lead to $M_1=335\pm0.5$ GeV and $M_{\tilde{u}_3}=1578\pm1$ GeV and a
precision of about 1 per mille, while our calculations would rather imply $M_1=330\pm2$
GeV and $M_{\tilde{u}_3}=1578\pm4$ GeV, i.e.\ a shift in $M_1$ by 5 GeV
or 1.5\% with an uncertainty of $0.5\%$, five times bigger than naively believed.

\section{Conclusion}
\label{Conclusion}

In summary, we have presented in this paper the first estimate of the theoretical uncertainty
of the SUSY dark matter relic density from renormalization scheme and scale variations.
Using three typical benchmark scenarios of a pMSSM with eleven free parameters, we have
analyzed in particular gaugino (co)annihilation into heavy quarks, gaugino-stop coannihilation
into top quarks and electroweak gauge and Higgs bosons or gluons, and stop-antistop annihilation
processes. Due to different renormalization schemes in particular in the top quark sector,
we have obtained results that differ from standard dark matter programs such as \MO\ already at
the tree level. We have quantified the impact of the renormalization scheme in this context
in particular for neutralino-stop coannihilation into top quarks and the lightest, SM-like
Higgs boson. We have also explained in detail how a renormalization scale dependence enters
all calculations already at tree level through coupling constants, in particular the trilinear
coupling $A_t$ for electroweak or the strong coupling $\alpha_s$ for strong processes,
through the running bottom quark mass, in particular in the resonant Higgs funnel, and through
the scale-dependent squark mixing angles in $t$- and $u$-channel squark-exchange processes.

Depending on the considered subprocesses and their relative importance for the calculation of
the total relic density, the renormalization scale dependence can differ significantly, but
it is reduced in almost all cases when NLO SUSY-QCD corrections are included. This was true
despite the fact that $\alpha_s$ enters often for the first time at that order and could be
traced to a slow, subdominant running of the strong coupling at the high scales of ${\cal O}$
(1 TeV) considered here. For neutralino-stop coannihilation into top quarks and Higgs bosons
we could demonstrate an enhanced perturbative stability of our mixed renormalization scheme
over a $\overline{\rm DR}$ definition of the top quark mass with a significantly reduced
$K$-factor and scale dependence.

As in our previous work, the stop-antistop annihilation channel showed the largest corrections
due to Sommerfeld enhancement at the cosmologically important low relative stop velocities.
The resummation of potential gluon exchanges introduced an additional dependence on the Coulomb
scale, which we chose to lie close to the Bohr radius of the would-be stoponium. The scale
uncertainty was then of similar order (about $\pm20\%$) as in the perturbative region despite
the fact that the correction could amount to factors of 10 in the low-velocity regime.

The net effect of our calculations is that the relic density cannot always be determined
theoretically with a precision of two percent similar to the experimental one (cf.\ Eq.\
\eqref{Planck}). Higher-order SUSY-QCD corrections rather induce important shifts of up to
50\% as we observed in our second benchmark scenario with important Sommerfeld enhancement
effects in stop-antistop annihilation (cf.\ Fig.\ \ref{fig:relicScenB}). The theoretical
uncertainty could now be more reliably estimated to be about six times as large as the
experimental one. In our
first benchmark scenario, dominated by gaugino (co)annihilation,
the relic density corrections reached only up to 10\%, and the NLO theory uncertainty became
comparable to the experimental one (cf.\ Fig.\ \ref{fig:RelicScenA}). An intermediate case
with important neutralino-stop coannihilation and additional other contributions was presented in
our third scenario (cf.\ Fig.\ \ref{fig:RelicScenC}). There the relic density corrections reached
almost 30\% and the theoretical uncertainty was reduced by almost a factor of two from LO to NLO,
but it remained about six times larger than the experimental one.
If one wanted to extract the pMSSM parameters from the relic density measurement, one would
consequently 
have to contend with shifts and uncertainties at the few percent
rather than the per mille level  in these parameters,
that one would naively extract from standard contour plots.

In this paper, we have of course only studied SUSY-QCD effects and ignored scheme and scale
uncertainties from the electroweak sector. While they are implicitly present to some extent
in the renormalization group running of our physical parameters, we leave their explicit
study for future work.

\acknowledgments

We thank M.~Meinecke for his contributions in the early stages of this work and A.~Pukhov for
providing us with the necessary functions to implement our results into the {\MO} code. Work
in M\"unster was partially supported by the Helmholtz Alliance for Astroparticle Physics (HAP),
by the DFG Graduiertenkolleg 2149, and by the DFG through grant No.\ KL 1266/5-1.
The work of J.H.\ has been performed within the Labex ILP (reference
ANR-10-LABX-63) as part of the Idex SUPER and received financial state aid
managed by the Agence Nationale de la Recherche as part of the
programme {\it Investissements d'Avenir} under the reference ANR-11-IDEX-0004-02.
J.H.\ would further like to acknowledge University College London, where an
early part of the work was performed and supported by the {\it London Centre for
Terauniverse Studies}, using funding from the European Research Council
via the Advanced Investigator Grant 267352. All figures have been produced
using {\tt Matplotlib} \cite{Hunter:2007}.


\bibliographystyle{apsrev}

\end{document}